%% file: pdf4eft.tex
\documentclass[prl,twoside,preprintnumbers,superscriptaddress,twocolumn,nofootinbib,nobibnotes]{revtex4}
\usepackage{amsmath,graphicx,slashed,multirow,color,ulem}
\usepackage{graphicx,epsfig,amssymb,bm,slashed,wasysym,multirow}
\usepackage[utf8]{inputenc}
\def \beq{\begin{equation}}
\def \eeq{\end{equation}}
\def \beqa{\begin{eqnarray}}
\def \eeqa{\end{eqnarray}}

\usepackage{ragged2e}

\usepackage{multirow,booktabs}

\usepackage{graphicx}
\usepackage{url}
\usepackage{cancel}
\usepackage{placeins}
\usepackage{bm}
\usepackage{hyperref}
\usepackage{amsmath}
\usepackage{amssymb}
\usepackage{listings}
\usepackage[toc, page]{appendix}
\usepackage{color}

\def\sw{s_W}
\def\cw{c_W}
\def\u{u(x,Q^2)}

\def\ub{\bar{u}(x,Q^2)}

\newcommand{\kz}{K_Z}

\usepackage{tabularx}
\newcolumntype{C}[1]{>{\centering\arraybackslash}p{#1}}

\def\bea{\begin{eqnarray}}
\def\eea{\end{eqnarray}}

\def\gsim{\mathrel{\rlap{\lower4pt\hbox{\hskip1pt$\sim$}}
    \raise1pt\hbox{$>$}}}         
\def\lsim{\mathrel{\rlap{\lower4pt\hbox{\hskip1pt$\sim$}}
    \raise1pt\hbox{$<$}}}         

\preprint{Nikhef 2019-009, DAMTP-2019-19, CP3-19-25, TIF-UNIMI-2019-5}

\newcommand{\be}{\begin{equation}}
\newcommand{\ee}{\end{equation}}
\newcommand{\bi}{\begin{itemize}}
\newcommand{\ei}{\end{itemize}}
\newcommand{\ben}{\begin{enumerate}}
\newcommand{\een}{\end{enumerate}}
\newcommand{\la}{\left\langle}
\newcommand{\ra}{\right\rangle}
\newcommand{\lc}{\left[}
\newcommand{\rc}{\right]}
\newcommand{\lp}{\left(}
\newcommand{\rp}{\right)}

\begin{document}
\title{Can New Physics hide inside the proton?}

\author{Stefano Carrazza}
\affiliation{Tif Lab, Dipartimento di Fisica, Universit\`a di Milano and
  INFN, Sezione di Milano, Via Celoria 16, I-20133 Milano, Italy}

\author{Celine Degrande}
\affiliation{Centre for Cosmology,
Particle Physics and Phenomenology (CP3),
Universit\'e Catholique de Louvain,
B-1348 Louvain-la-Neuve, Belgium }

\author{Shayan Iranipour }
\affiliation{DAMTP, University of Cambridge, Wilberforce Road, Cambridge, CB3 0WA, United Kingdom }

\author{Juan Rojo}
\affiliation{Department of Physics and Astronomy, VU University Amsterdam, De Boelelaan 1081, NL-1081, HV Amsterdam, The Netherlands}
\affiliation{Nikhef, Science Park 105, NL-1098 XG Amsterdam, The Netherlands}

\author{Maria Ubiali }
\email{m.ubiali@damtp.cam.ac.uk }
\affiliation{DAMTP, University of Cambridge, Wilberforce Road, Cambridge, CB3 0WA, United Kingdom }

\date{\today}

\begin{abstract}
Modern global analyses of the 
structure of the proton include collider measurements which probe energies
well above the electroweak scale.
While these provide powerful constraints
on the parton distribution functions (PDFs),
they are also sensitive to
beyond the Standard Model (BSM) dynamics if these affect
the fitted distributions.
Here we present a first simultaneous determination
of the PDFs and BSM effects
from deep-inelastic structure function data by means of the NNPDF framework.
We consider representative four-fermion operators from the
SM Effective Field Theory (SMEFT), quantify to which extent their effects
modify the fitted PDFs, and assess
how the resulting bounds on the SMEFT degrees
of freedom are modified.
Our results demonstrate how BSM effects that might otherwise be reabsorbed into the PDFs
can be systematically disentangled.
\end{abstract}

\maketitle


Searches for physics beyond the Standard Model (BSM) at high-energy
colliders can be divided
into two main categories: direct searches, 
aiming to detect the production of new heavy particles,
and  indirect searches, 
whose goal is to identify subtle deviations in the  interactions
and properties of the SM
particles.
The latter would arise from virtual quantum effects involving
BSM dynamics at energies well beyond the collider centre-of-mass
energy.
Both strategies are actively pursued at the LHC by 
exploiting its unique energy reach~\cite{Rappoccio:2018qxp}
and its thriving  program of
precision
measurements~\cite{Dawson:2018dcd,Cristinziani:2016vif,Erler:2019hds}.

In this context, the SM Effective Field Theory (SMEFT)~\cite{Weinberg:1978kz,Buchmuller:1985jz,Arzt:1994gp,Grzadkowski:2003tf, AguilarSaavedra:2008zc,Nomura:2009tw, AguilarSaavedra:2010zi,Grzadkowski:2010es,Brivio:2017vri} represents a 
powerful model-independent approach to identify, interpret, and correlate
potential BSM effects from precision measurements under the assumption
that the new physics scale, $\Lambda$, is
well above the energies probed by the experimental data.
Here, BSM effects
can be parametrised at low energies
in terms of dimension-six operators, $\mathcal{O}_i$,  constructed
from SM fields satisfying its symmetries:
\begin{equation}
  \label{eq:SMEFTdef}
\mathcal{L}_{\rm SMEFT} = \mathcal{L}_{\rm SM} +\sum_{i=1}^{N} \frac{a_i}{\Lambda^{2}} \mathcal{O}_i \, ,
\end{equation}
where $\mathcal{L}_{\rm SM}$ is the SM Lagrangian, $\{a_i\}$ are
the Wilson coefficients parametrising the high-energy BSM dynamics, and $N$ is the number
of non-redundant operators.
These Wilson coefficients can be constrained
from measurements ranging from Higgs, gauge boson, and electroweak
precision observables~\cite{Ellis:2018gqa,Biekotter:2018rhp,Almeida:2018cld,deBlas:2016ojx} to top
quark production~\cite{Buckley:2015nca,Hartland:2019bjb},
flavour observables~\cite{Aebischer:2018iyb,Straub:2018kue},
and low-energy processes~\cite{Falkowski:2017pss}, among others.
In the case of LHC data, high-energy processes such as Drell-Yan, diboson, and
top-quark production at large invariant
masses~\cite{Farina:2016rws,Greljo:2017vvb,Alioli:2017nzr,Alioli:2018ljm,Maltoni:2019aot,Alte:2017pme,Farina:2018lqo}
play a key role since energy-growing
effects often enhance the sensitivity to the SMEFT contributions.

Several of the high-energy LHC measurements that constrain
the SMEFT parameter space are also used to provide stringent constraints
on the proton's parton distribution functions (PDFs)~\cite{Butterworth:2015oua,Gao:2017yyd}.
Prominent examples include the
large-$x$ gluon from top-quark pair~\cite{Guzzi:2014wia,Czakon:2016olj} and jet
production~\cite{Harland-Lang:2017ytb,Rojo:2014kta}, and the quark-flavor
separation from high-mass Drell-Yan and $W$ and $Z$ boson production
in association with jets~\cite{Boughezal:2017nla,Malik:2013kba,Giuli:2017oii,ATL-PHYS-PUB-2019-016}.
This implies that  BSM effects, if present in the high-energy tails
of those distributions, could end up being ``fitted away'' into the PDFs.
These concerns are particularly acute
for the full exploitation of the Run II and III datasets, as well
as from the High-Luminosity phase~\cite{Azzi:2019yne} where
many PDF-sensitive observables will reach the few-TeV region~\cite{Khalek:2018mdn}.

In this work, we want to address two main questions.
First, how can one assess whether BSM effects have been absorbed into
the fitted PDFs?
And second, how are the bounds on the SMEFT coefficients modified
if the PDFs used as input to determine them had been fitted using a consistent
BSM theory?
To answer them, we present here a first simultaneous determination of the proton's PDFs and
the SMEFT Wilson coefficients $\{a_i\}$ by means of the
NNPDF framework~\cite{Ball:2014uwa,Ball:2012cx,Ball:2008by,Ball:2010de}.
As a proof of concept, we consider the constraints from deep-inelastic scattering (DIS)
structure functions on representative four-fermion operators.
This way, we are able
to quantify to which extent SMEFT effects (which parametrise
general BSM dynamics) can be reabsorbed into the flexible
neural-network based PDF parametrisation~\cite{Rojo:2018qdd}.
This has required extending the NNPDF framework 
such that cross-sections can be evaluated including
BSM corrections at the fit level.
We also assess how the bounds on the SMEFT coefficients
are modified in this joint fit as compared to the traditional approach
where PDFs are kept fixed.
See~\cite{Aaron:2011mv,Abramowicz:2019uti} for
related {\tt xFitter}~\cite{Alekhin:2014irh} studies
restricted to H1 and ZEUS data and to one-parameter fits.

Here we study the impact of operators of the form
 \be
 \mathcal{O}_{lq} = \left(\bar{l}_R\gamma^\mu l_R\right)\left(\bar{q}_R\gamma_\mu q_R\right) \label{eq:operatorsDef} \, ,\,\,q=u,d,s,c \, ,
\ee
where $l_R$ and $q_R$ stand for right-handed charged leptons
and quarks fields. 
We assume coupling universality in
the lepton sector but not in the quark one,
in order to evade the strong constraints
from LEP precision data~\cite{Han:2004az}.
These operators lead to a energy-growing
effects and their contributions are weighted by the corresponding
PDFs, two properties that provide powerful handles for discriminating them.

The calculation of the SMEFT corrections from the $\mathcal{O}_{lq}$ operators
in Eq.~(\ref{eq:operatorsDef})
to DIS
structure functions can be performed in analogy with the
corresponding SM computation.
For instance, $F_2$ will now contain terms linear and quadratic in $a_u$,
the coefficient of $\mathcal{O}_{lu}$ in Eq.~(\ref{eq:SMEFTdef}):
\bea
\Delta F_2^{\rm smeft} &\supset& \frac{x}{12 e^4}\Bigg( 4 a_{u} e^2 \frac{Q^2}{\Lambda^2}(1+4\kz\sw^4)+ 3a_{u}^2 \frac{Q^4}{\Lambda^4}\Bigg) \nonumber
  \\
  && \,\,\times\Big( \u + \ub \Big), \label{eq:SF_SMEFT}
  \eea
  where $\kz = Q^2 /(4\cw^2\sw^2(Q^2+M_Z^2))$,
  $\sw=\sin\theta_W$, $\cw=\cos\theta_W$, and $u$~($\bar{u}$)
  represents the up (anti-)quark PDF.
  The terms linear in $a_u$ 
  arise from the interference 
  with the SM amplitudes and are suppressed as $Q^2/\Lambda^2$.
  Similar expressions to  can be evaluated for the contributions
  from $\mathcal{O}_{ld}$, $\mathcal{O}_{ls}$, and $\mathcal{O}_{lc}$,
  and for the parity-violating structure function $xF_3$,
  while $\Delta F_L^{\rm smeft}=0$ at leading order.
  In this work we will keep only the leading $\mathcal{O}\lp \Lambda^{-2}\rp $ terms
  in Eq.~(\ref{eq:SF_SMEFT}), 
  though we have verified that results are stable upon the addition of the
  $\mathcal{O}\lp \Lambda^{-4}\rp$ ones.
  These SMEFT-augmented structure functions
  have been implemented
  into {\tt APFEL}~\cite{Bertone:2013vaa,Bertone:2016lga}.
  The DGLAP equations for the scale
  evolution of the PDFs are unaffected.
  
\begin{figure}[t]
  \begin{center}
    \makebox{\includegraphics[width=0.99\columnwidth]{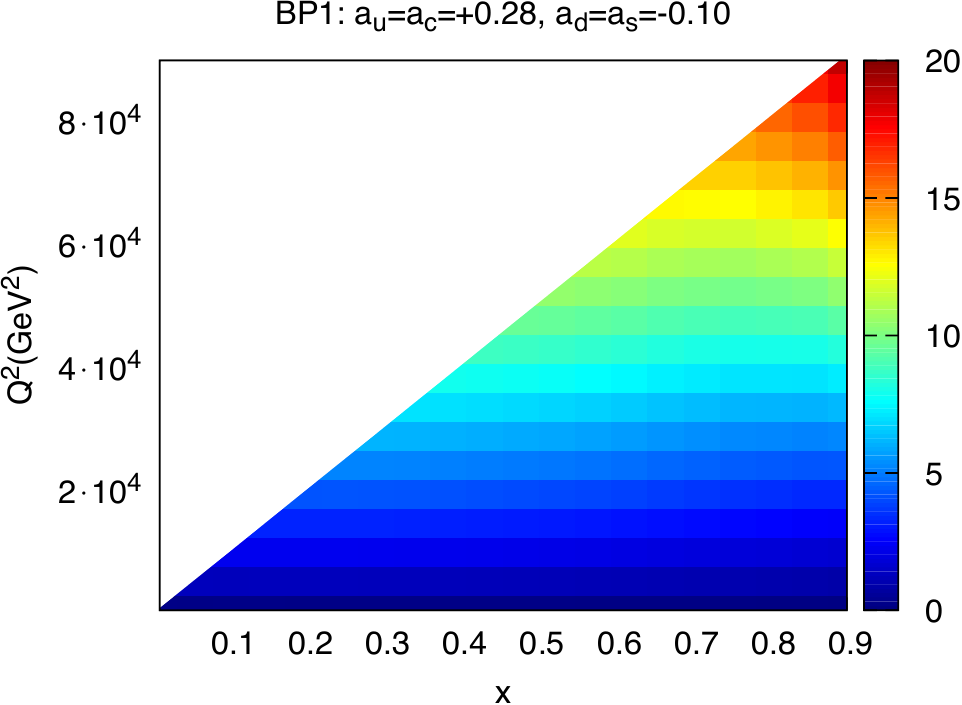}}
   \end{center}
  \vspace{-0.6cm}
  \caption{The percentage SMEFT-induced
    shift, Eq.~(\ref{eq:SMEFTshifts}),
    for the $e^-p$ neutral current DIS cross-section
    at $\mathcal{O}\lp Q^2/\Lambda^{2}\rp$ for a representative choice
    of $\{a_q\}$ as a function of $x$ and $Q^2$.
  }
  \label{fig:SMEFT-corrections-BP}
\end{figure}

Since SMEFT effects are suppressed as $Q^2/\Lambda^2$, only measurements
involving large momentum transfers $Q^2$ will be sensitive to them.
The only DIS experiment that has explored the region $Q \gsim M_W$ is
HERA~\cite{Klein:2008di},
whose legacy structure function data~\cite{Abramowicz:2015mha}
reach up to $Q_{\rm max} \simeq 250$ GeV.
In Fig.~\ref{fig:SMEFT-corrections-BP} we
display the percentage shift in the $e^-p$ neutral
current (NC) DIS cross-section, 
\be
\label{eq:SMEFTshifts}
\Delta_{\rm smeft}\equiv  \lp d^2\sigma^{{\rm NC}}/dxdQ^2\rp\Big/\lp d^2\sigma_{\rm SM}^{{\rm NC}}/dxdQ^2\rp -1
\,,
\ee
as a function of $x$ and $Q^2$,
for a representative choice
of  coefficients given by
$a_u=a_c=0.28$ and $a_d=a_s=-0.10$.
As in the rest of the paper,  we assume here that $\Lambda=1$ TeV.
The corrections depend only mildly on Bjorken-$x$ and increase
rapidly with $Q$, reaching up to $\simeq 20\%$ for
the upper HERA kinematic limit.

Given that for a sizable region of the SMEFT parameter space the shifts
Eq.~(\ref{eq:SMEFTshifts}) are comparable or bigger than the experimental
uncertainties of the precise HERA structure functions, the latter can be
exploited to
impose bounds on the allowed ranges of the  coefficients $\{a_q\}$.
First of all, we evaluate the values of $\chi^2_{\rm tot}(\{a_q\})$ for
the DIS measurements used in  NNPDF3.1~\cite{Ball:2017nwa},
corresponding to $n_{\rm dat}=3092$ data points
from BCDMS, SLAC, NMC, CHORUS,
NuTeV, and HERA.
In this calculation, we use
NNPDF3.1 NNLO DIS-only as input with
consistent theory settings such as FONLL-C~\cite{Forte:2010ta}
and fitted charm~\cite{Ball:2016neh}.
This is repeated for a range
of SMEFT benchmark points (BPs) (listed in the
Appendix) and for the  $N_{\rm rep}=100$
Monte Carlo replicas.
The resulting  $\chi^2_{\rm tot}(\{a_q\})$ values are then fitted to a quadratic form,
\be
\label{eq:hessian}
\chi^2_{\rm min} + \sum_{q,q'=u,d,s,c} H_{qq'}\lp a_q-a_q^{\rm (min)}\rp
\lp a_{q'}-a_{q'}^{\rm (min)}\rp
\ee
where $H_{qq'}$ are the elements of the Hessian matrix in the quark
flavour space.
Note that Eq.~(\ref{eq:hessian}) is exact if the
$\mathcal{O}(\Lambda^{-4})$ corrections are neglected,
else it is valid only close to a local minimum.
We have performed the fits of the SMEFT coefficients both
varying a single operator at a time (individual fits)
as well as varying the four of them simultaneously  and then marginalizing over
each one.

In Table~\ref{tab:bounds-prefit} we
indicate the 90\% confidence level (CL) intervals
for the four coefficients obtained
with fixed input PDFs. 
We compare the individual bounds
with the marginalised ones from the four-dimensional fits,
without and with PDF uncertainties.
In the former case, theory calculations are obtained using
the central replica.
In the latter case, we compute the bounds for the $N_{\rm rep}=100$
replicas and take the envelope of the 90\% narrower ones.

\input{bounds-prefit.tex}

The most stringent bounds
are obtained for $a_u$, followed by $a_d$, and then $a_c$ and $a_s$.
This is consistent with
the fact that the SMEFT corrections proportional
to $a_q$ are weighted by the corresponding PDFs in Eq.~(\ref{eq:SF_SMEFT}),
and that in the HERA region $u(x) \gsim d(x) \gg s(x),\,c(x)$.
The marginalised bounds are looser than the individual ones by up to
an order of magnitude, highlighting the relevance of exploring simultaneously
the widest possible region of the parameter space.
PDF uncertainties turn out to be moderate.
For the individual fits, the bounds are stable
upon the inclusion of $\mathcal{O}\lp Q^4/\Lambda^4\rp$ terms.

The main limitation of the bounds reported in Table~\ref{tab:bounds-prefit}
is that they are affected by double counting, since the same HERA data
was already included in the NNPDF3.1 fit
used here to evaluate the DIS structure functions
with SMEFT effects.
The very same problem arises for the interpretation of collider
measurements that are used to constrain  both the PDFs and
the SMEFT parameter space, such as jet, Drell-Yan, and top quark pair production.
To bypass this limitation, the way forward is provided by
the simultaneous extraction of the PDFs and the SMEFT degrees of freedom
$\{a_q\}$, in the same way as in joint extractions
of PDFs and the strong coupling constant~\cite{Ball:2018iqk}.

\begin{figure}[t]
  \begin{center}
    \makebox{\includegraphics[width=0.99\columnwidth]{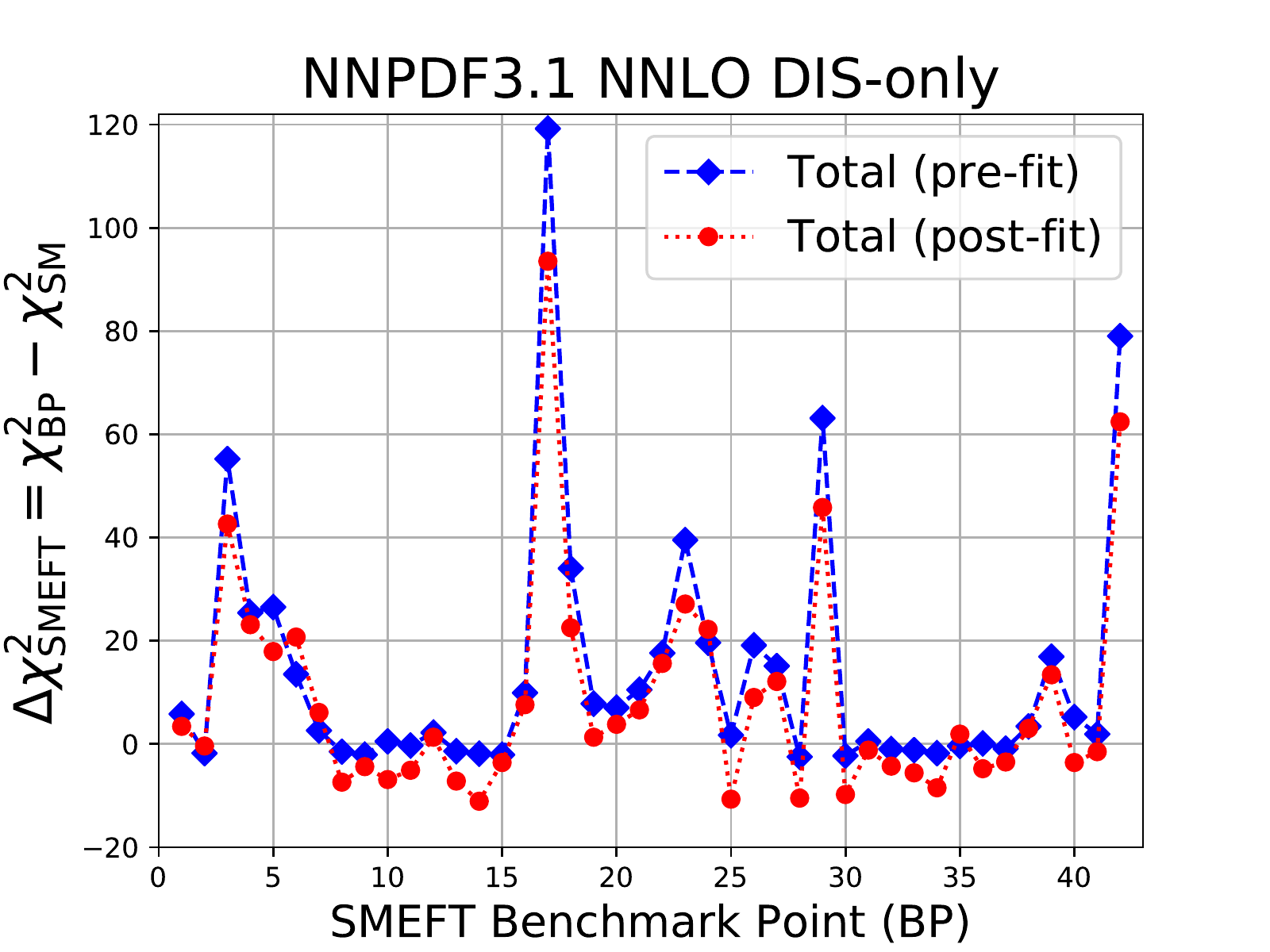}}
   \end{center}
  \vspace{-0.5cm}
  \caption{The difference in  $\chi^2_{\rm tot}$
    with respect to the SM in the fits 
    with different BPs, compared to the fixed-PDF
    results.
  }
  \label{fig:deltachi2}
\end{figure}

We have thus carried out variants of the  NNPDF3.1 NNLO DIS-only fit
now using as theory input the structure functions with SMEFT corrections.
These fits have been performed for 
the same BPs as in the fixed-PDF analysis, and are based on $300$
replicas to tame statistical fluctuations.
Defining $\Delta \chi^2_{\rm smeft}=\chi^2_{\rm tot}-\chi^{2(\rm SM)}_{\rm tot}$,
we find that the BP with the largest improvement (deterioration)
with respect to the SM has $\Delta\chi^2_{\rm smeft} \simeq -10$
($\simeq 90$), see Fig.~\ref{fig:deltachi2}.
In all cases, $\chi^2_{\rm tot}$ decreases as compared
to the pre-fit (fixed-PDF) result, indicating that SMEFT
effects are being partially reabsorbed into the PDFs.

From Fig.~\ref{fig:deltachi2} one expects that 
in the fits with SMEFT corrections
the resulting PDFs will be distorted as compared to their SM-based
counterparts.
Here the flexible NNPDF parametrisation is suitable
to robustly assess to what extent such effects can be reabsorbed into the PDFs.
Firstly,
we find that the  quark valence distributions
are rather similar to those of the SM case, see the Appendix.
The reason is that quark PDFs are dominantly fixed by the
moderate $Q^2$ fixed-target
DIS data, and thus unaffected by the high-$Q^2$
HERA structure functions.

\begin{figure}[t]
  \begin{center}
    \makebox{\includegraphics[width=0.99\columnwidth]{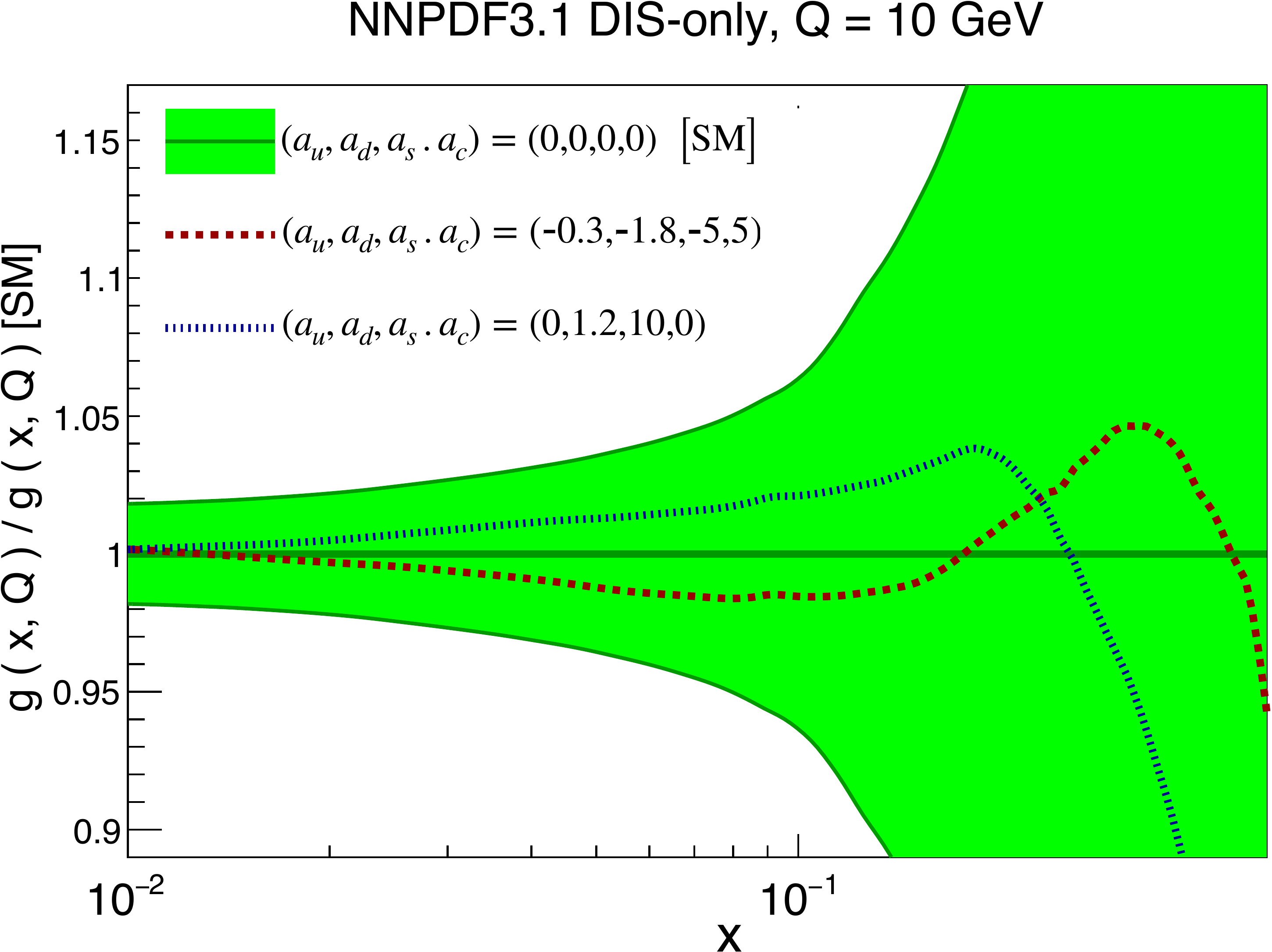}}
  \end{center}
  \vspace{-0.55cm}
  \caption{The gluon PDF in the fits
    with two representative SMEFT BPs
    (for $\Lambda=1$ TeV),
    normalised to the SM result.
  }
  \label{fig:xpdf-smeft}
\end{figure}

More significant differences are observed for the gluon PDF.
Within a DIS-only fit, the gluon is
mostly constrained  from the scaling violations between the low- and high-$Q^2$ data,
which 
are strongly modified in the presence of
energy-growing SMEFT effects.
In Fig.~\ref{fig:xpdf-smeft} we show
the gluon in the fits based on the
$(a_u,a_d,a_s,a_c)=(-0.3,-1.8,-5,5)$
and $(0,1.2,10,0)$ BPs, normalised to the SM
and where PDF uncertainties are only displayed for the latter.
These are two of the BPs leading to the largest
deviations from the SM  at the $\chi^2$ level, with
$\Delta \chi^2_{\rm smeft}\simeq65$ and 41 at the pre-fit level respectively,
while also being consistent
with the bounds from the HERA data in Table~\ref{tab:bounds-prefit}.
We find that the SMEFT-induced distortions can be comparable with the
PDF uncertainties and thus should be taken into account.
These distortions would be even more pronounced in a global
fit, where the gluon can be extracted with higher precision.

\begin{figure}[t]
  \begin{center}
    \makebox{\includegraphics[width=0.99\columnwidth]{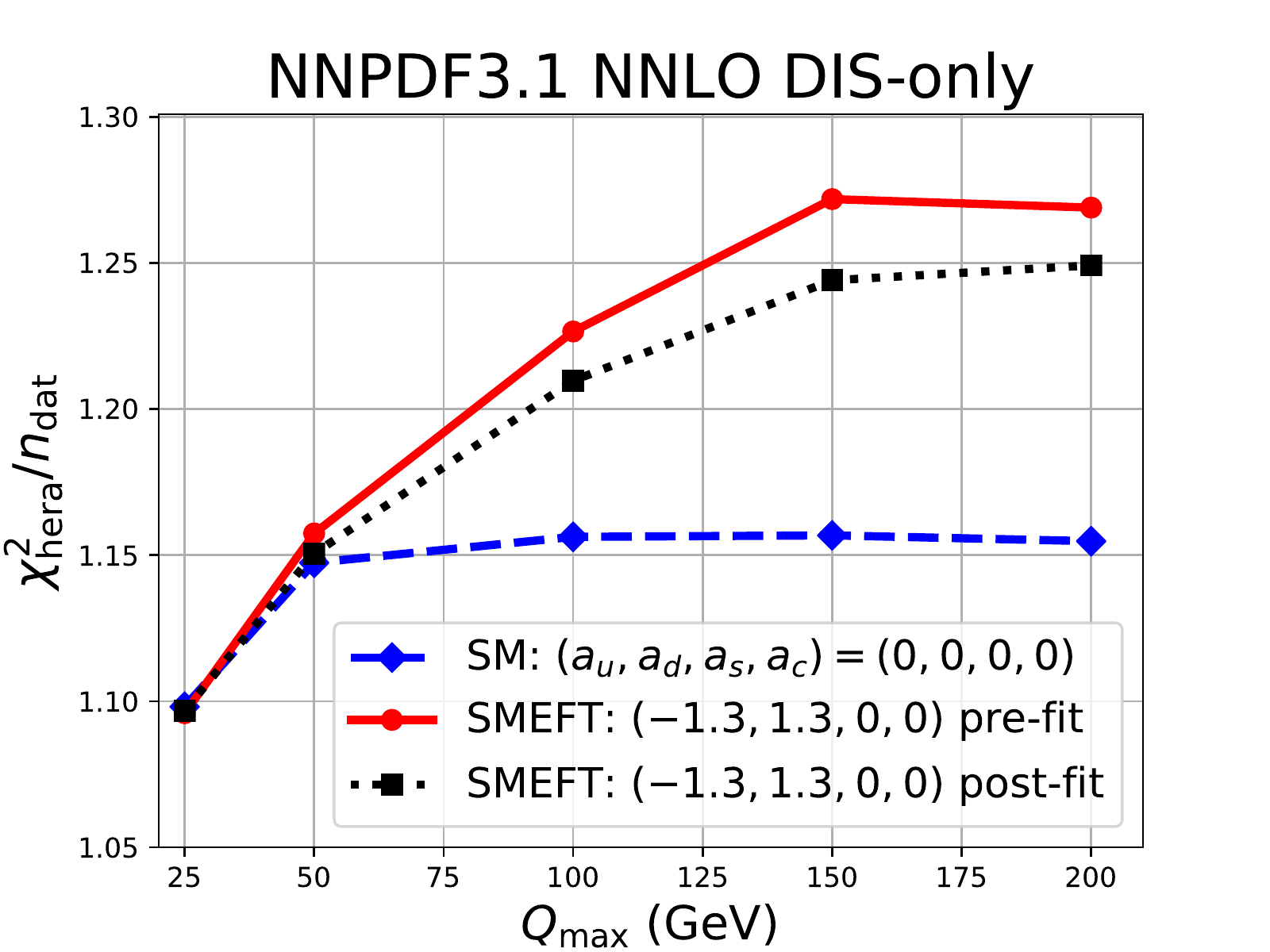}}
   \end{center}
  \vspace{-0.55cm}
  \caption{The dependence
    of  $\chi^2_{\rm hera}/n_{\rm dat}$ on $Q_{\rm max}$ in the case of the
    SM and for one representative SMEFT BP, for which
   we indicate both the pre-fit and the post-fit values.
  }
  \label{fig:chi2_vs_qmax}
\end{figure}

The different energy scaling of the SMEFT effects as compared to the
QCD ones (polynomial in the former, logarithmic 
in the latter) can be exploited to disentangle BSM dynamics from
QCD ones within the PDF fit.
In Fig.~\ref{fig:chi2_vs_qmax} we display
 $\chi^2_{\rm hera}/n_{\rm dat}$
as a function of the cut $Q_{\rm max}$
that fixes the
maximum value that enters the $\chi^2$ evaluation.
Results are shown both in the SM
and in the SMEFT for $a_u=a_d=-1.3$ and $a_s=a_c=0$,
and in the latter case
both for the pre-fit
(fixed-PDF) and post-fit cases.
While for $Q_{\rm max}\gsim 50$ GeV the value of $\chi^2_{\rm hera}/n_{\rm dat}$ 
is flat for the SM case, there is a rapid degradation
in the fit quality for the SMEFT case.
This result further highlights that BSM effects cannot be
completely ``fitted away''.
Such distinctive trend in the high-energy behaviour of the theory
would represent a smoking gun for BSM effects,
similar to how BFKL dynamics were recently identified from
small-$x$ HERA data~\cite{Ball:2017otu}.

In Table~\ref{tab:bounds-postfit} we indicate
the individual and marginalised 90\% CL intervals for the SMEFT coefficients
from this joint extraction together with the PDFs, see
Table~\ref{tab:bounds-prefit} for
the fixed-PDF ones.
We find that the bounds are rather similar in both cases, consistent
with the evidence from Figs.~\ref{fig:deltachi2}-\ref{fig:chi2_vs_qmax} that
SMEFT corrections are only partially reabsorbed in the PDFs.
As expected, the individual limits are somewhat broader at the post-fit level.
The marginalised bounds are affected by a sizable statistical uncertainty
associated with the finite number of replicas.
The latter is estimated by Gaussianly fluctuating the $\chi^2_{\rm tot}$ values
of each BP around their central values by their bootstrap uncertainty,
and keeping only those fluctuations leaving a positive-definite Hessian.
The resulting distribution of fit minima, eigenvalues, and eigenvectors
are used to estimate these statistical errors, finding
in particular that they are larger than the central value associated
to
the smallest eigenvalue and therefore that a flat direction,
mainly in the  $(a_s-a_c)$ plane, could not be excluded.

\input{bounds-postfit.tex}

\begin{figure}[t]
  \begin{center}
    \makebox{\includegraphics[width=0.95\columnwidth]{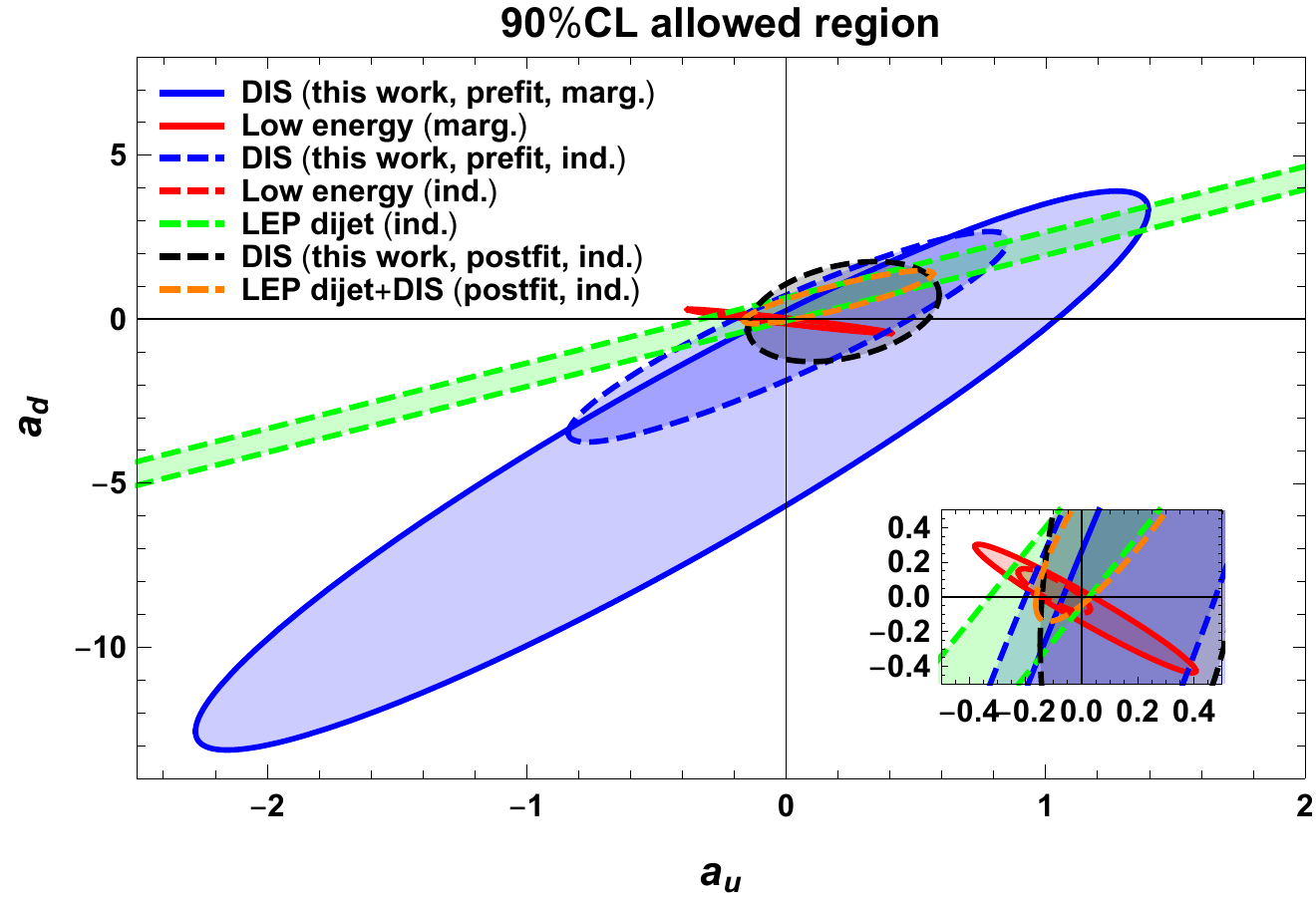}}
   \end{center}
  \vspace{-0.55cm}
  \caption{The 90\% CL marginalised and individual bounds in the $\lp a_u, a_d\rp$ plane
    from this work compared to those from~\cite{Falkowski:2017pss} (dijet
    and parity)
    and to the individual bound from dijet data.
  }
  \label{fig:bounds-comparison-lowenergy}
\end{figure}

Other studies have quantified the constraints on four-fermion
operators such as those of Eq.~(\ref{eq:operatorsDef}), and a
compilation of the information from precision LEP data
and low-energy measurements was presented in~\cite{Falkowski:2017pss}.
In Fig.~\ref{fig:bounds-comparison-lowenergy} we compare
the 90\% CL bounds in the $\lp a_u, a_d\rp$ plane
from our work (both pre-fit and post-fit level) 
with those from both LEP dijet data and parity measurements.
We also show the individual bounds from the former since,
contrary to the parity data, these are independent
on the modeling of the nucleon structure.
We find that our precise bound for $a_u$ is comparable to previous studies,
while those for $a_d$, $a_s$, and $a_c$ are less competitive.
This encouraging result emphasizes
the potential of high-energy collider data for the simultaneous
extraction of both PDFs and SMEFT degrees of freedom.

To summarise, in this work we have systematically analysed
the interplay between PDF and SMEFT fits, using the HERA structure functions that provide the backbone
of all modern PDF extractions as a case study.
Our results represent the successful proof-of-concept of a
program aiming to exhaustively disentangle potential BSM effects
in high-energy measurements
that might otherwise be reabsorbed into the PDFs.
The next steps in this program will be to extend our study
to a global dataset, including LHC data, and to a wider operator basis.

\paragraph{Acknowledgements.}
We are grateful to A. Falkowski and  K. Mimouni
for discussions concerning~\cite{Falkowski:2017pss}. We thank F.
Maltoni, E. Mereghetti, T. You, B. Allanach and the SUSY working group
for insightful discussions.
S.~C. is supported by the European Research Council under
the European Union's Horizon 2020 research and innovation Programme
(grant agreement n.740006).
C.~D. is supported by by the Fund for Scientific Research F.N.R.S. through the F.6001.19 convention.
 J.~R. is supported by the European Research Council Starting
Grant ``PDF4BSM'' and by the Netherlands Organization for Scientific
Research (NWO).
M.~U. and S.~I. are partially supported by the STFC grant
ST/L000385/1 and the Royal
Society grant RGF/EA/180148.
M.~U. is funded by the
Royal Society grant and DH150088.

\input{pdf4eft.bbl}
\onecolumngrid
\appendix

\input{app-fits.tex}

\end{document}

%% file: bounds-prefit.tex
\begin{table}[h]
  \centering
  \footnotesize
   \renewcommand{\arraystretch}{1.30}
   \begin{tabular}{|C{1.3cm}|C{2cm}|C{2cm}|C{2cm}|}
      \hline
	   &   Individual   & \multicolumn{2}{c|}{Marginalised}    \\
	   &    &  no PDF unc   &  w PDF unc   \\
\hline
$a_u$  &   $\lc -0.1, +0.4 \rc$  &
	   $\lc -2.3, +1.4\rc$ &  $\lc -3.6, +2.7\rc$        \\
$a_d$  &   $\lc -1.6, +0.4\rc$  &
	   $\lc -13, +3.9\rc$   &  $\lc -19, +11  \rc$      \\
$a_s$  &   $\lc -2.8, +4.2 \rc$  &
	   $\lc -18, +29\rc$ &  $\lc -36, +47 \rc$       \\
$a_c$  &   $\lc -2.6, +1.2 \rc$  &
	   $\lc -13, +7.0\rc$ &  $\lc -21, +15  \rc$      \\
\hline
  \end{tabular}
   \caption{\small The 90\% CL intervals (for $\Lambda=$ 1 TeV)
     for the coefficients extracted with
     fixed PDFs, comparing 
     individual and marginalised bounds
     with and without PDF uncertainties.
     \label{tab:bounds-prefit}
  }
\end{table}

%% file: bounds-postfit.tex
\begin{table}[h]
  \centering
  \footnotesize
  \renewcommand{\arraystretch}{1.30}
  \begin{tabular}{|C{1.5cm}|C{2.3cm}|C{2.3cm}|}
            \hline
	   &   Individual   & Marginalised    \\
\midrule
\hline
$a_u$  &   $\lc 0.0, +0.5 \rc$  & $\lc -0.4, +2.4\rc $   \\
$a_d$  &   $\lc -1.1, +0.8\rc$  & $\lc -4.4, +4.5 \rc $  \\
$a_s$  &   $\lc -4.5, +3.6 \rc$ & $\lc -61, +39\rc $ \\
$a_c$  &   $\lc -2.4, +0.7 \rc$ & $\lc -29, +2.7\rc $   \\
\hline
\bottomrule
  \end{tabular}
  \caption{\small Same as Table~\ref{tab:bounds-prefit}
    for the simultaneous determination of the
     PDFs and the SMEFT coefficients.
	\label{tab:bounds-postfit} 
  }
\end{table}

%% file: app-fits.tex
\section{Mapping the SMEFT parameter space and impact on the PDF fits}

In this Appendix, we provide additional information about the 
fits based on theory calculations that include SMEFT corrections.
We describe how  we map the parameter space
associated to the SMEFT four-fermion operators of Eq.~(\ref{eq:operatorsDef}),
and in particular
present our choice of benchmark points (BPs).
We quantify systematically the impact on the PDFs of the SMEFT
corrections for each of the BPs, and also study in further
detail their mutual correlation.

First of all, in Table~\ref{tab:listfits} we list the BPs that
have been used to reconstruct the Hessian matrix in Eq.~(\ref{eq:hessian}),
both at the pre-fit (fixed input PDFs) and at the post-fit levels.
For each BP, we indicate the values
     of the total and HERA $\chi^2$, denoted by $\chi^2_{\rm tot}$
     and $\chi^2_{\rm hera}$ respectively, both in the case of a common fixed input
     PDF set (``pre-fit'' column) and once the PDFs have been fitted to the theory
     based on the corresponding BP (``post-fit'' column).
     Here $\chi^2_{\rm hera}$ contains only the contributions
from the neutral
and charged-current inclusive structure function data, but
not that from the heavy quark structure functions.
     The pre-fit values of the $\chi^2$ are obtained from the central replica
     of the NNPDF3.1 NNLO DIS-only set, while those of the post-fit case
     are computed from the average over the $N_{\rm rep}=300$ replicas
     available for each BP.
     Both at the pre-fit and the post-fit levels,
     we also indicate the absolute difference
     between $\chi^2_{\rm tot}$ in the SMEFT and in the SM, that is,
\be
\Delta \chi^2_{\rm smeft}=\chi^2_{\rm tot}-\chi^{2(\rm SM)}_{\rm tot} \, .
\ee
The total number of
data points in these fits is $n_{\rm dat}=3092$, while the number
of data points for the HERA inclusive structure function
data (both neutral current and charged current) is $n_{\rm dat}=1145$.
Recall that throughout this work we assume $\Lambda=1$ TeV.

\begin{figure}[h]
  \begin{center}
    \makebox{\includegraphics[width=0.55\columnwidth]{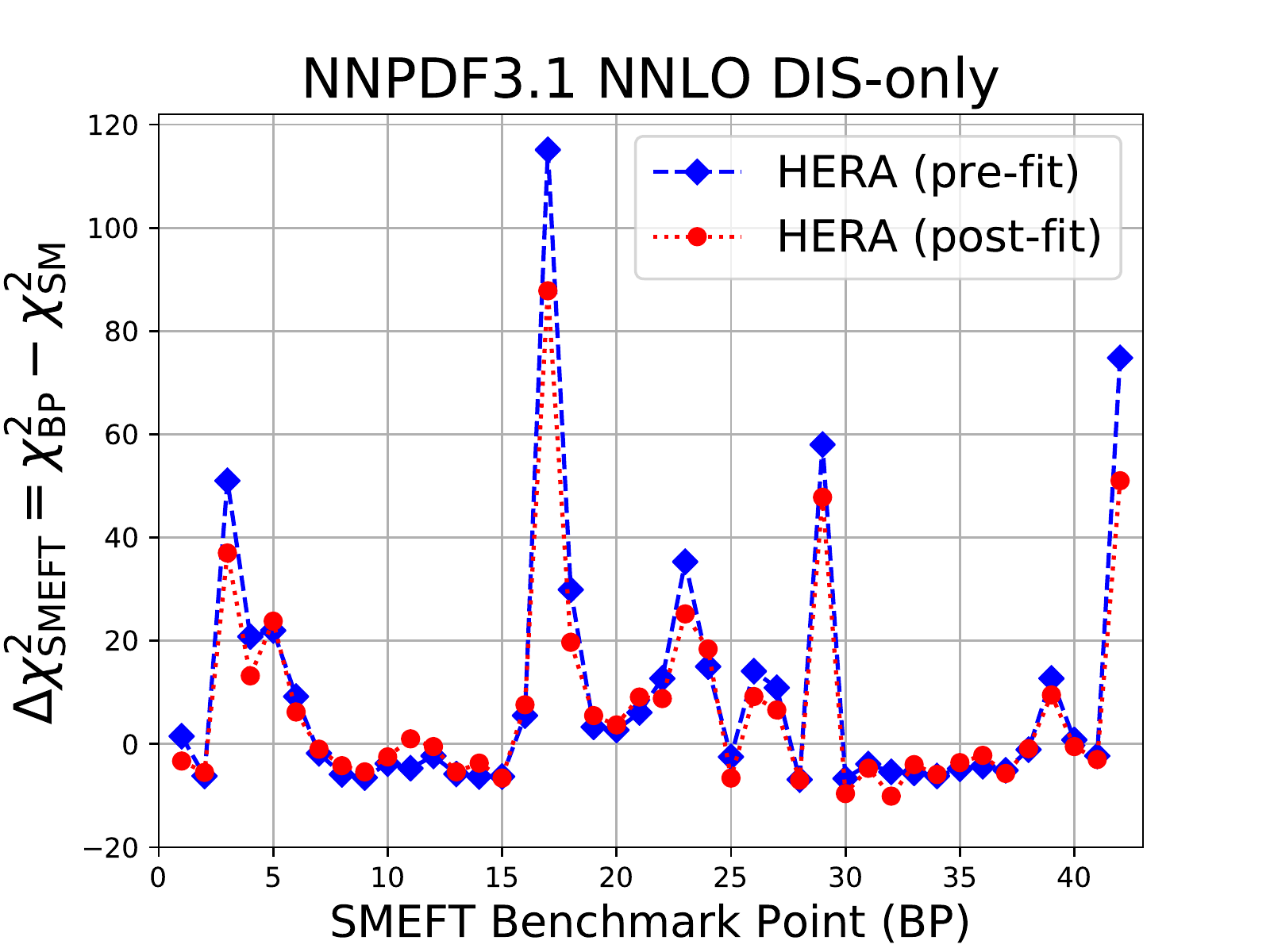}}
   \end{center}
  \vspace{-0.5cm}
  \caption{Same as Fig.~\ref{fig:deltachi2}, now for the contribution
    of the inclusive HERA structure functions.
  }
  \label{fig:deltachi2_hera}
\end{figure}

The graphical representation of the results for $\Delta \chi^2_{\rm smeft}$
based on the total dataset were displayed in Fig.~\ref{fig:deltachi2}.
In Fig.~\ref{fig:deltachi2_hera} we show the same comparison
for the values of $\Delta \chi^2_{\rm smeft}$ between the pre-fit and post-fit
cases, now restricted to the contribution from the HERA structure functions.
The impact of the SMEFT effects in the global PDF fit is clearly seen
to be driven by the
contributions from the
 HERA structure function data, and in particular for the large-$Q^2$ bins.
As in the case of the total dataset, we observe how the values
of $\Delta\chi^2_{\rm smeft}$ decrease in the post-fit case as
compared to the pre-fit one, due to SMEFT corrections
being partially reabsorbed into the fitted PDFs.

\input{chi2-smeft-nnpdf31dis.tex}

From Table~\ref{tab:listfits} it is possible to identify for
which BPs the effects of including the SMEFT
corrections is more significant, as indicated by the largest variations
in $\Delta \chi^2_{\rm smeft}$, and for which these effects
can be neglected.
The largest variations are found for BP17, defined by $a_u=-a_d=-1.3$
and $a_s=a_c=0$, for which $\Delta \chi^2_{\rm smeft}\simeq 121~(95)$
at the pre-fit~(post-fit) level.
For some BPs the values of the $\chi^2$ are actually improved
once SMEFT corrections are included, such as
for BP25, defined with $a_u=0.3$, $a_d=1.2$, and $a_s=a_c-5.0$, for
which $\Delta \chi^2_{\rm smeft}$ varies from $+3.5$ at the pre-fit
level to $\simeq -9$ at the post-fit level.
Note, however that some of the BPs for which $\Delta \chi^2_{\rm smeft}$ is largest,
such as BP3, BP17, or BP42, are excluded at the 90\% CL as indicated
in Tables~\ref{tab:bounds-prefit} (pre-fit) and~\ref{tab:bounds-postfit} (post-fit).

Next we summarise the main results of the fits listed in
Table~\ref{tab:listfits} at the PDF level.
The differences between the PDFs from the fits
    based on SM theory and those that include SMEFT
    corrections can be quantified by means of 
the PDF pull, defined for example in the case of the gluon as follows:
\be
\label{eq:PDFpull}
P_g(x,Q^2) = \frac{\la g(x,Q^2) \ra\Large|_{\rm smeft} - \la g(x,Q^2) \ra\Large|_{\rm SM} }{\sigma_g(x,Q^2)|_{\rm SM}} \, ,
\ee
and likewise for other flavours.
In Eq.~(\ref{eq:PDFpull}),
$\la g(x,Q^2) \ra$ indicates the fit central value
(average over replicas) for a given SMEFT BP and in the SM, while $\sigma_g(x,Q^2)$
corresponds to the standard deviation for the fit based on SM theory.
In other words, the pull measures the distortion in the PDF central value induced by the SMEFT
corrections in units of the 68\% CL SM uncertainty.
A value of the pull $|P| \ge 1$ would indicate an SMEFT-induced
distortion which is not contained within the corresponding PDF 68\%
confidence level uncertainty band.

\begin{figure}[t]
  \begin{center}
    \includegraphics[width=0.49\textwidth]{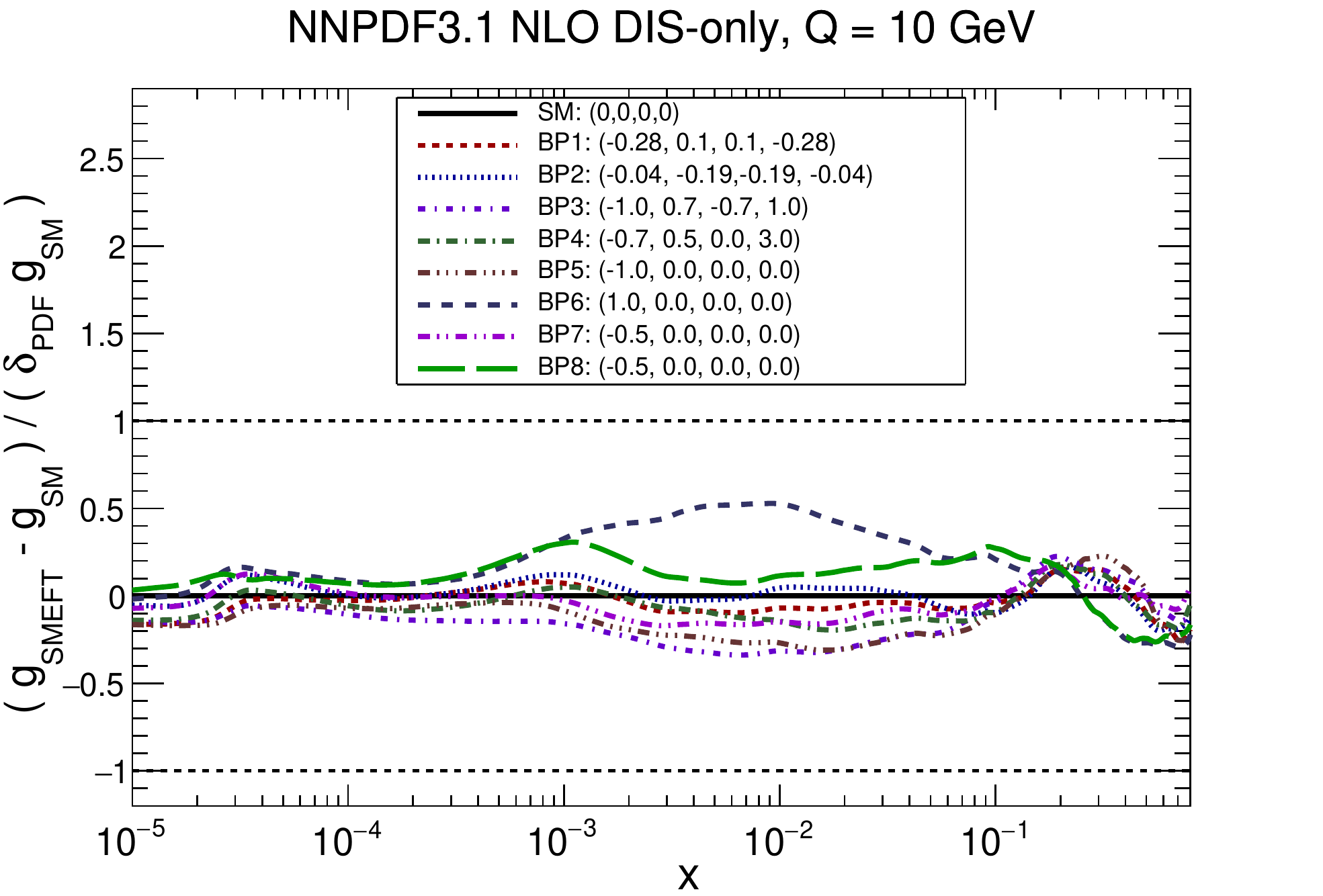}
    \includegraphics[width=0.49\textwidth]{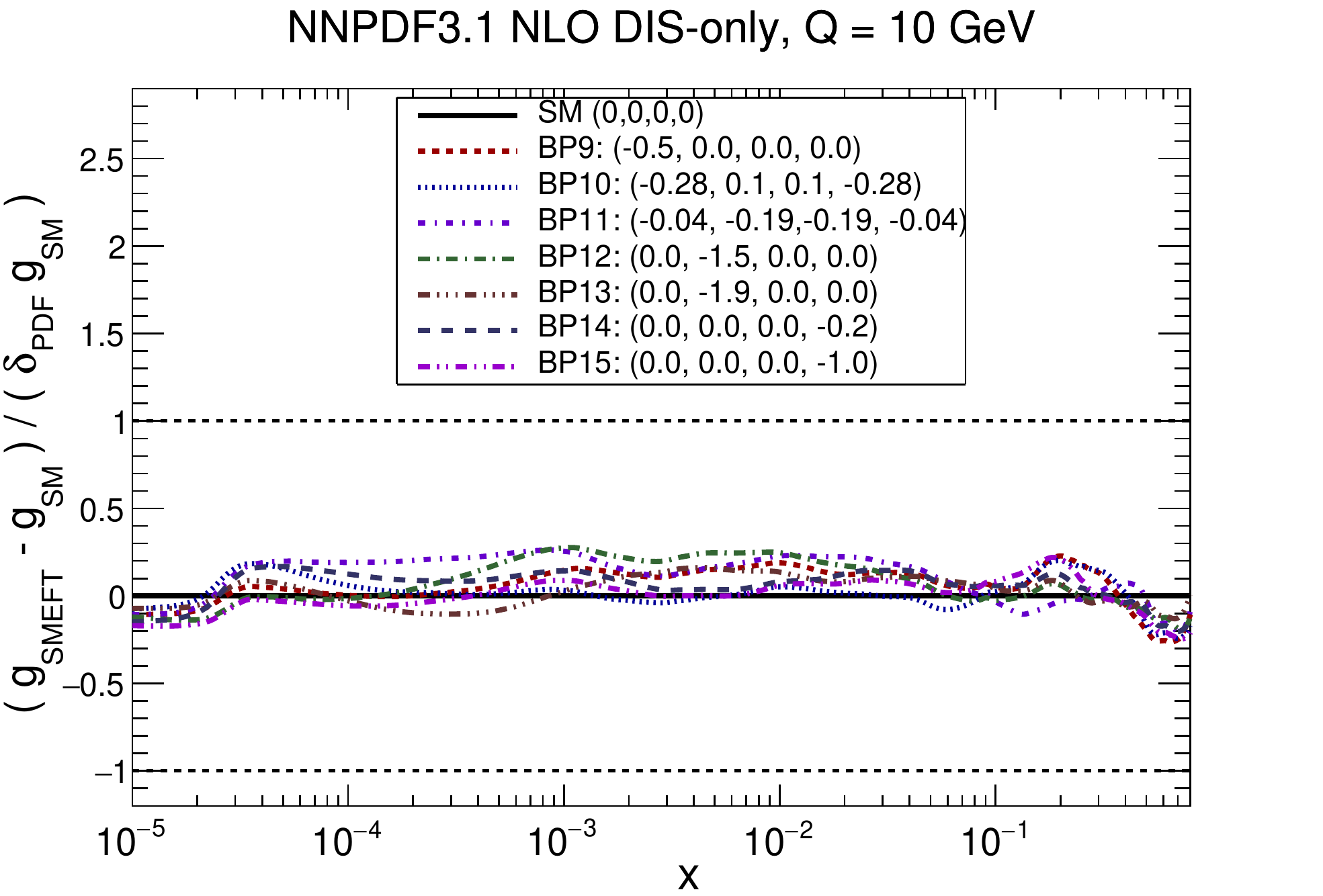}
    \includegraphics[width=0.49\textwidth]{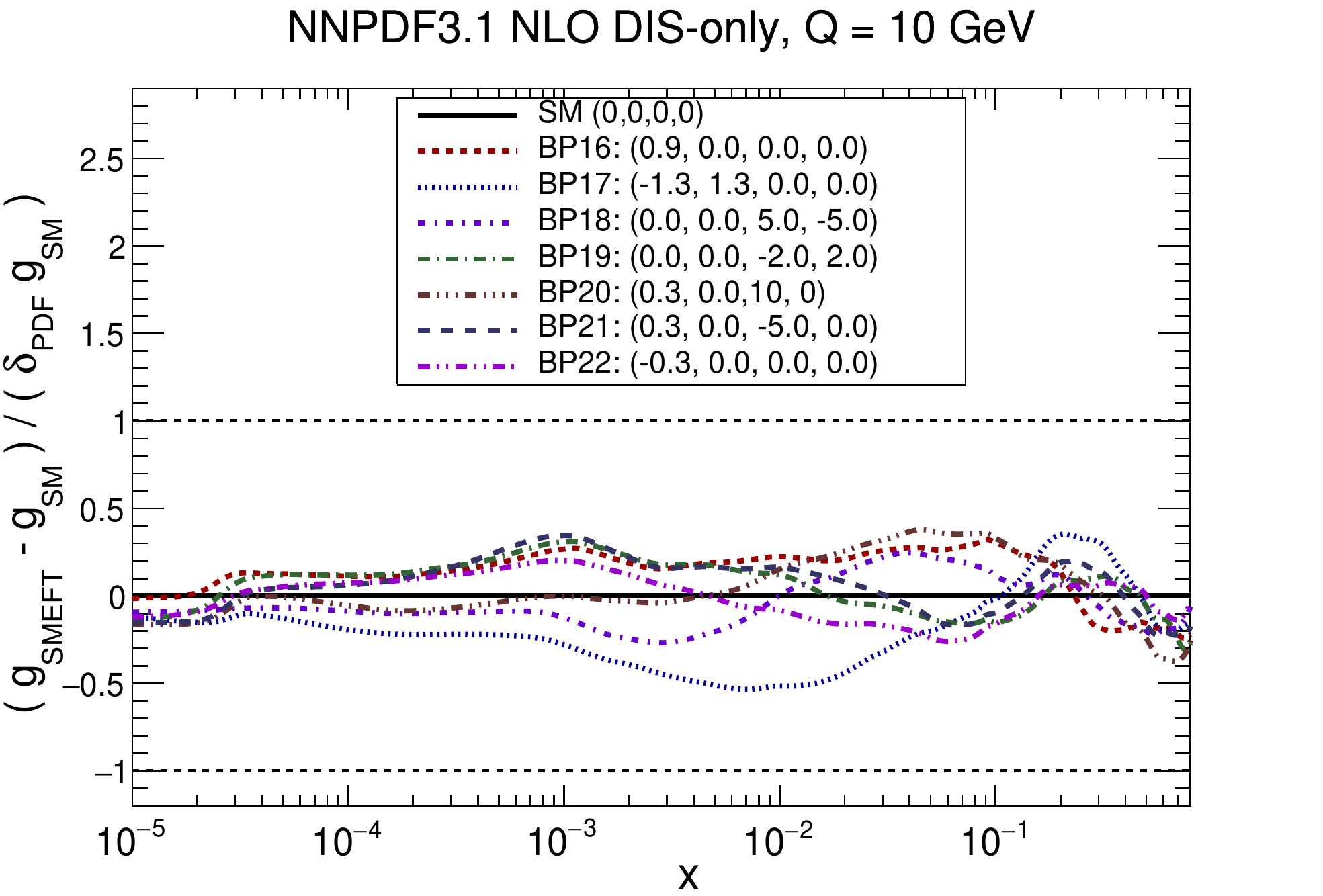}
    \includegraphics[width=0.49\textwidth]{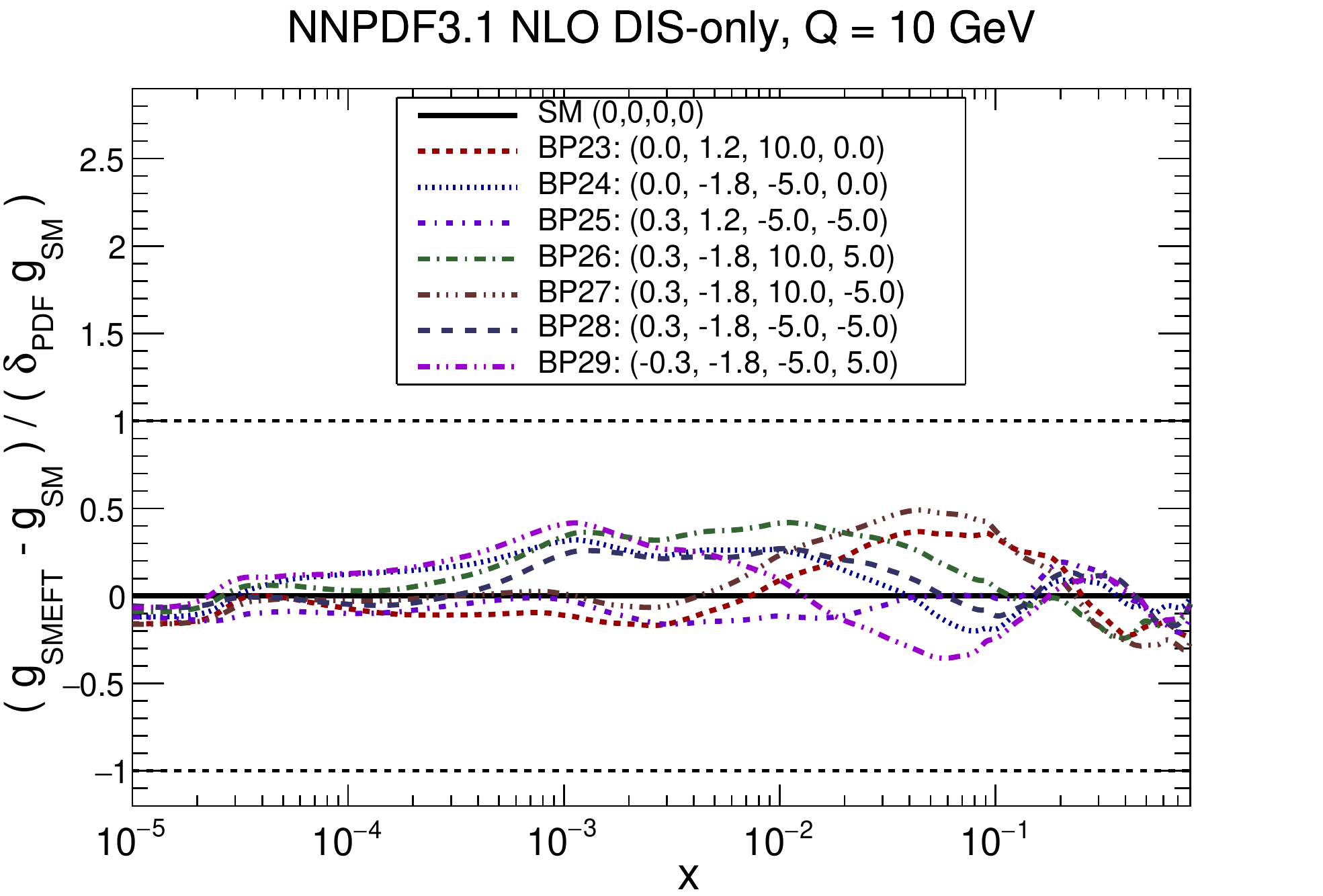}
    \includegraphics[width=0.49\textwidth]{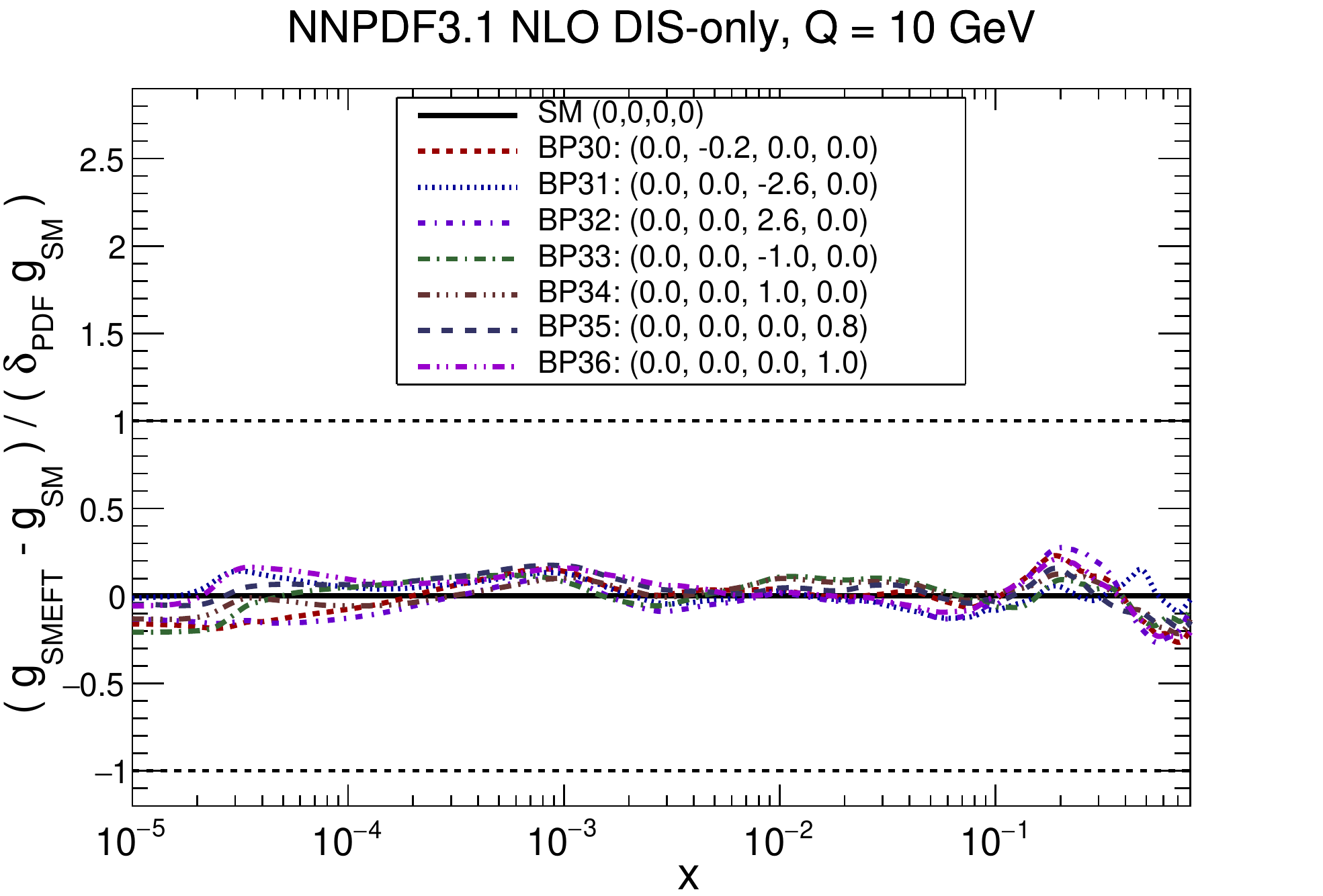}
    \includegraphics[width=0.49\textwidth]{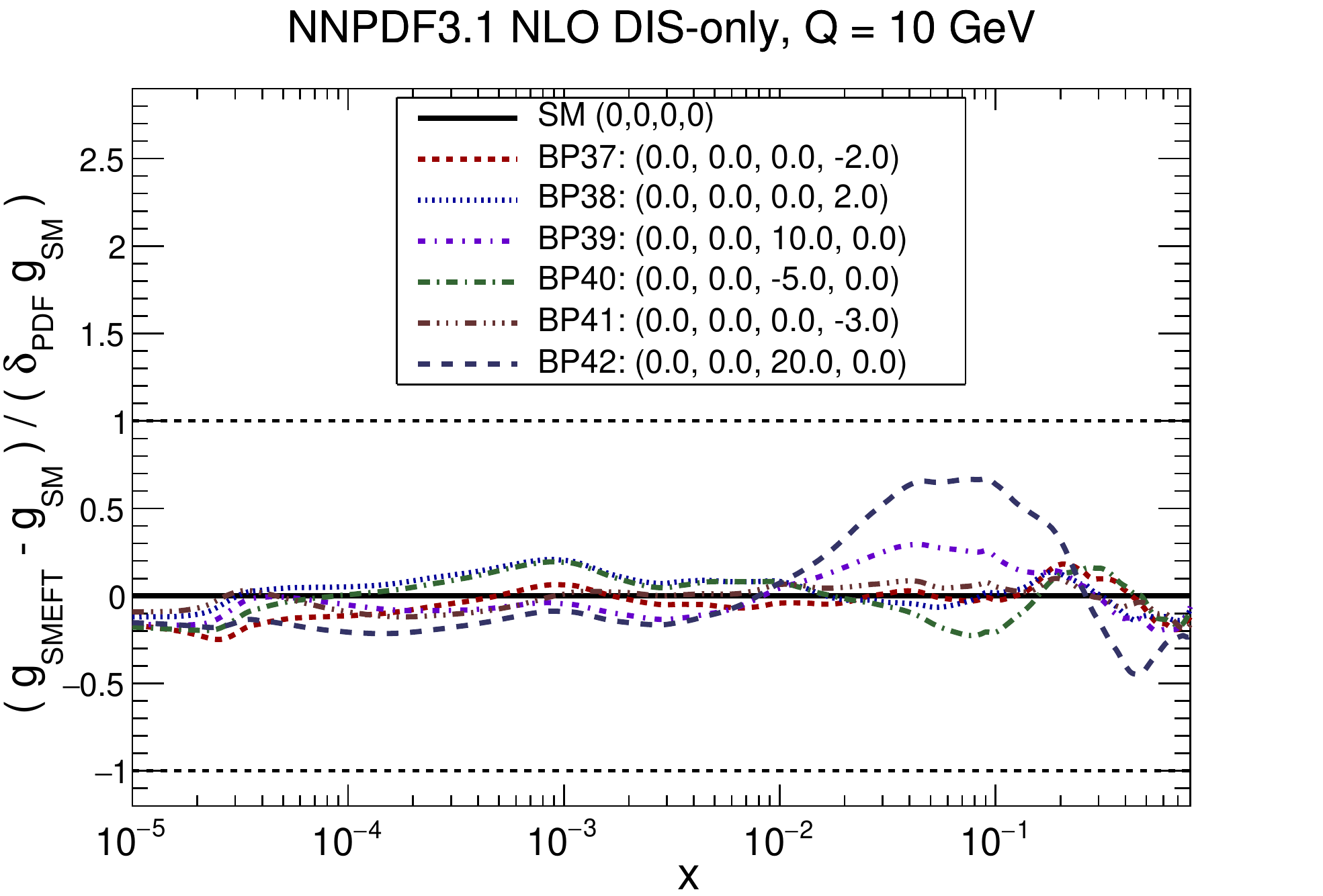}
  \end{center}
  \vspace{-0.3cm}
  \caption{The pull, Eq.~(\ref{eq:PDFpull}), for the gluon PDF between the fits
    based on the SM theory and those that include SMEFT
    corrections.
    The comparison is performed at $Q=10$ GeV for
    the benchmark points listed in Table~\ref{tab:listfits},
    where in each case we indicate the values of the $\lp a_u,a_d,a_s,a_d\rp$
    coefficients.
  }
  \label{fig:smeft-pull}
\end{figure}

\begin{figure}[t]
  \begin{center}
    \includegraphics[width=0.49\textwidth]{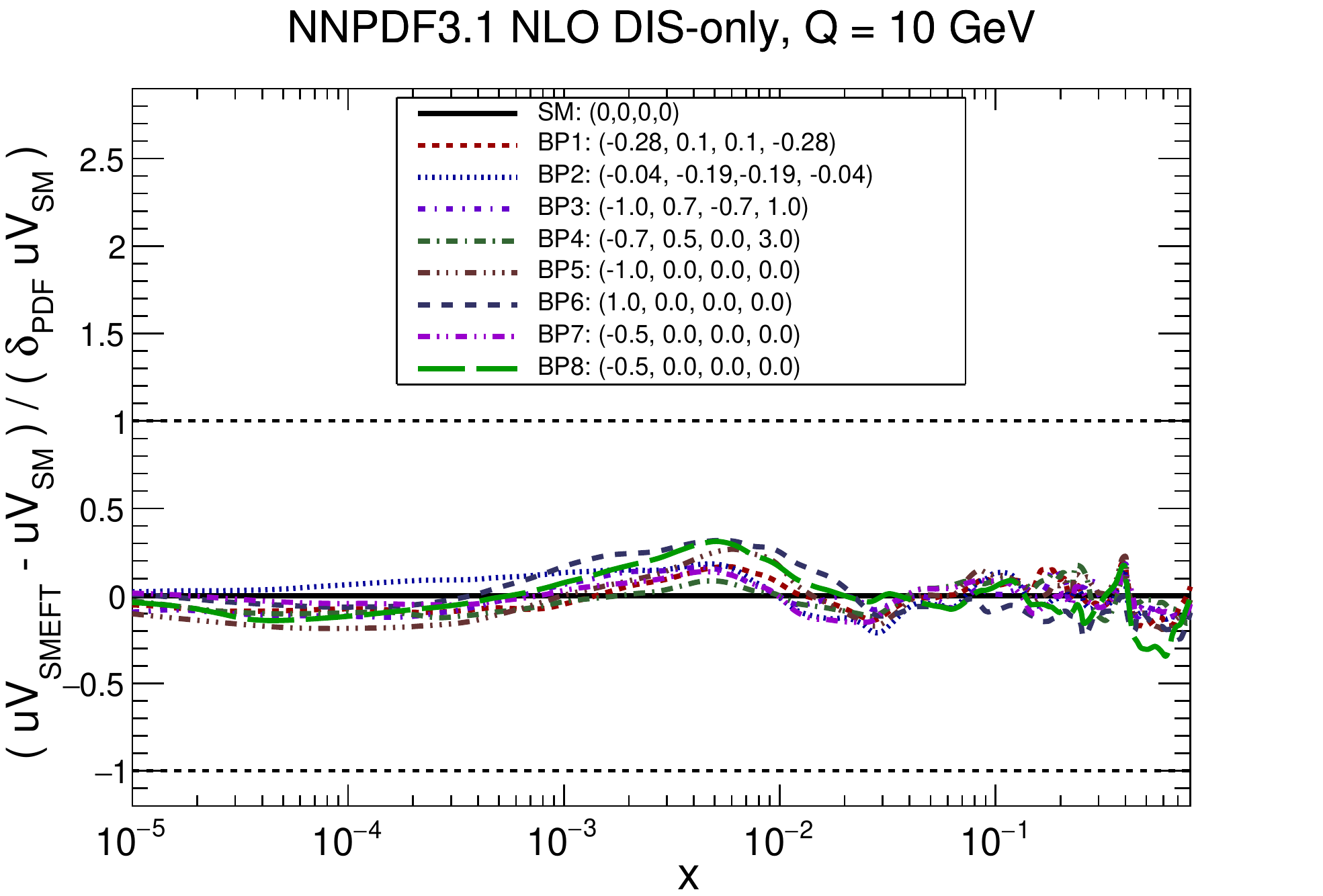}
    \includegraphics[width=0.49\textwidth]{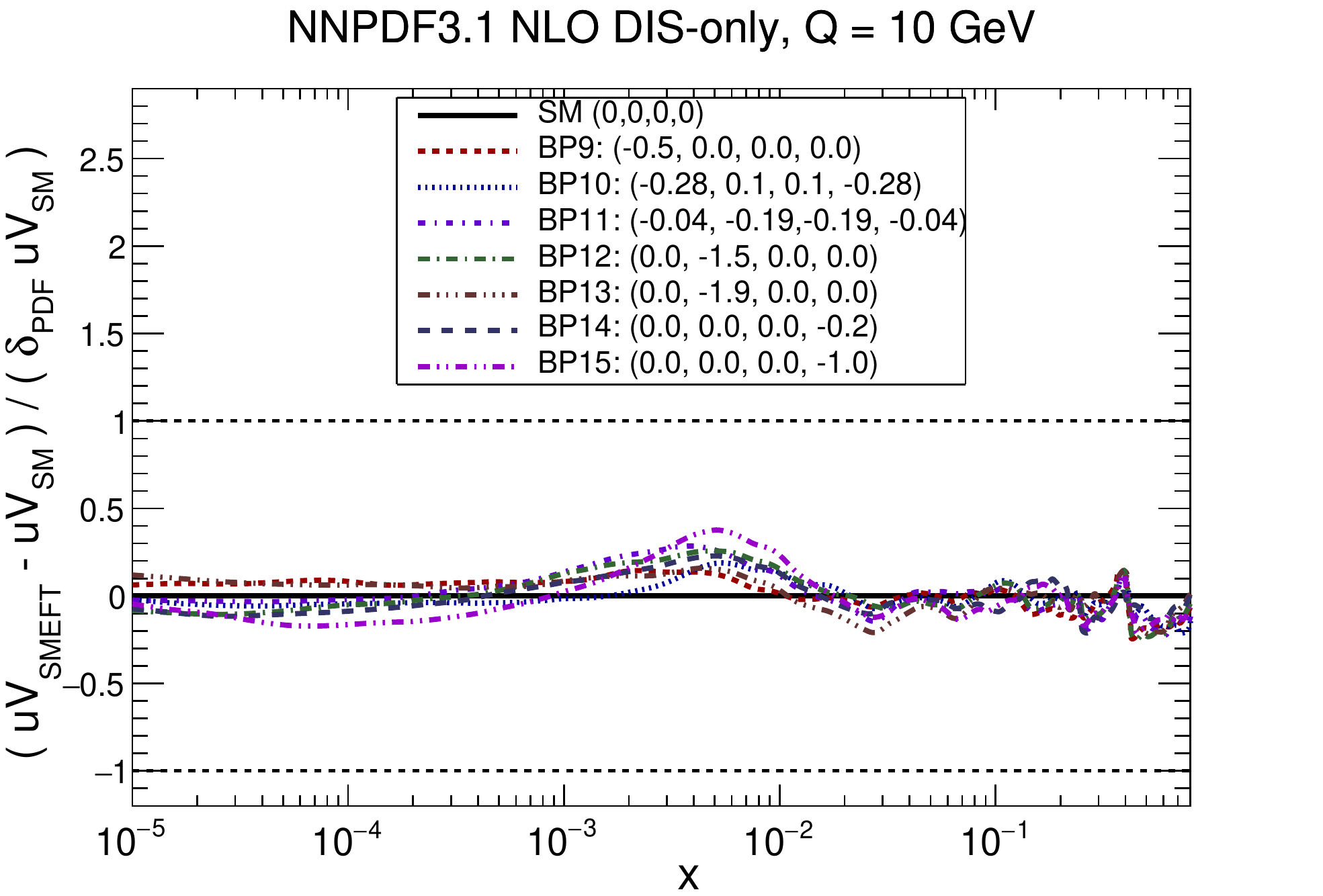}
    \includegraphics[width=0.49\textwidth]{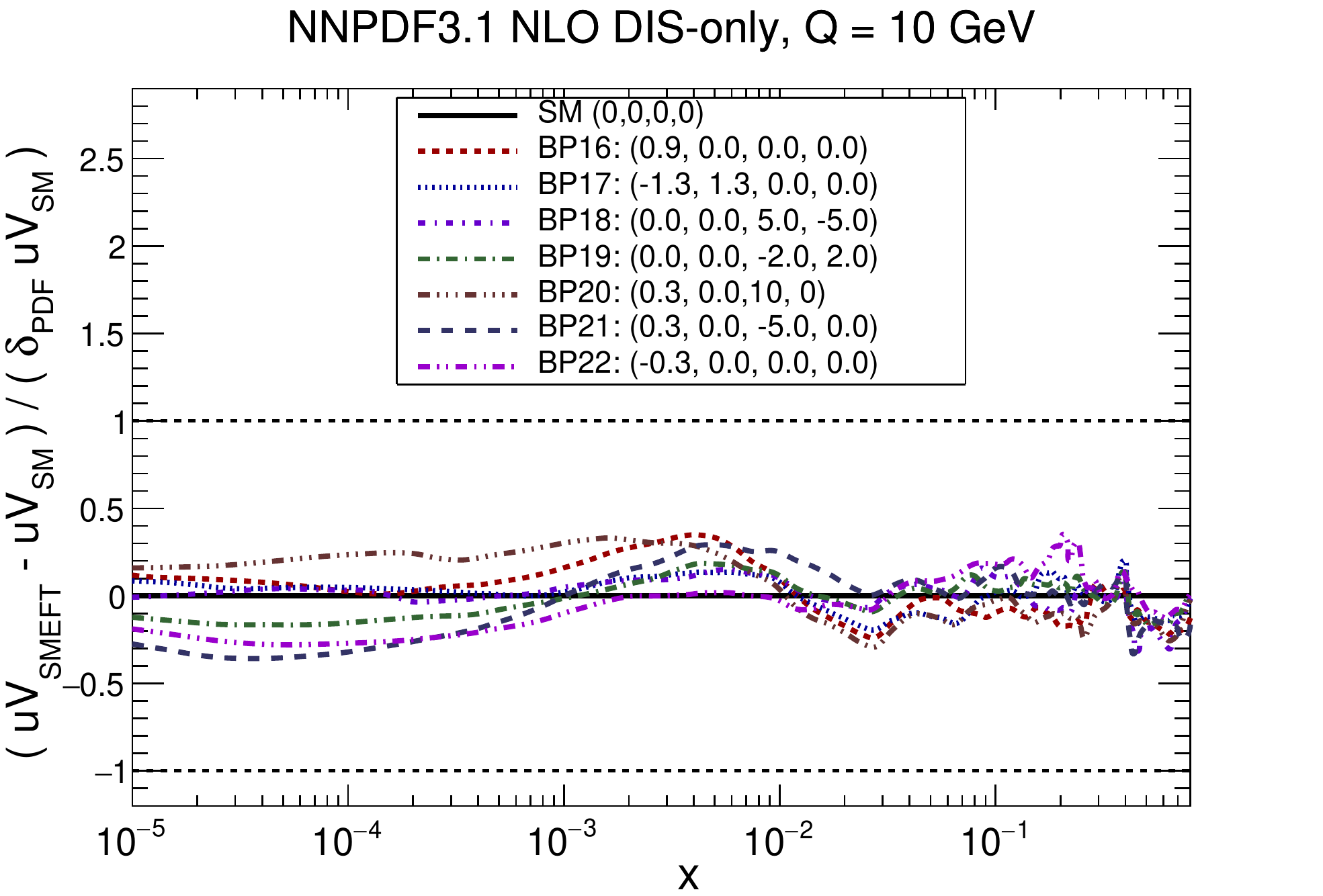}
    \includegraphics[width=0.49\textwidth]{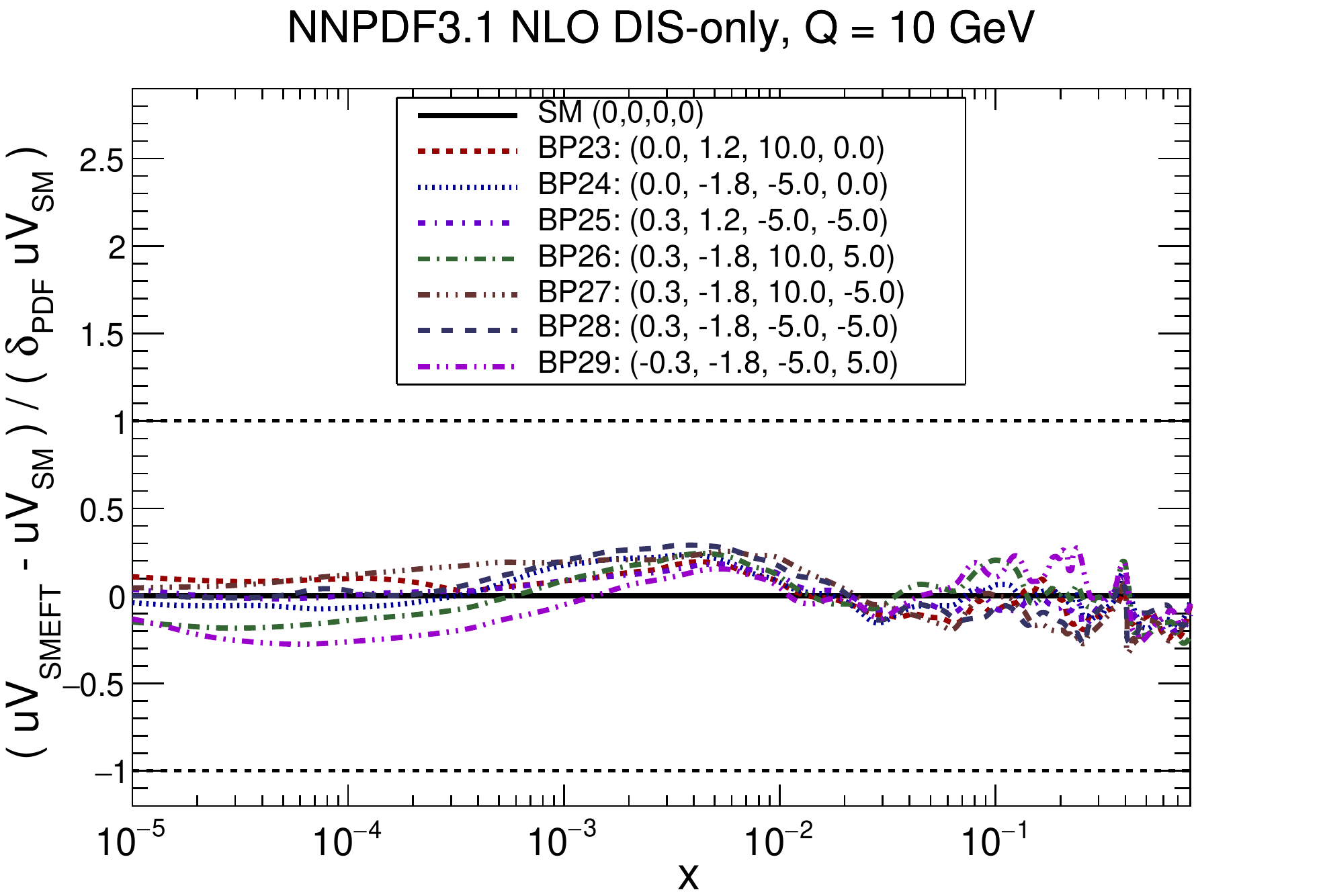}
    \includegraphics[width=0.49\textwidth]{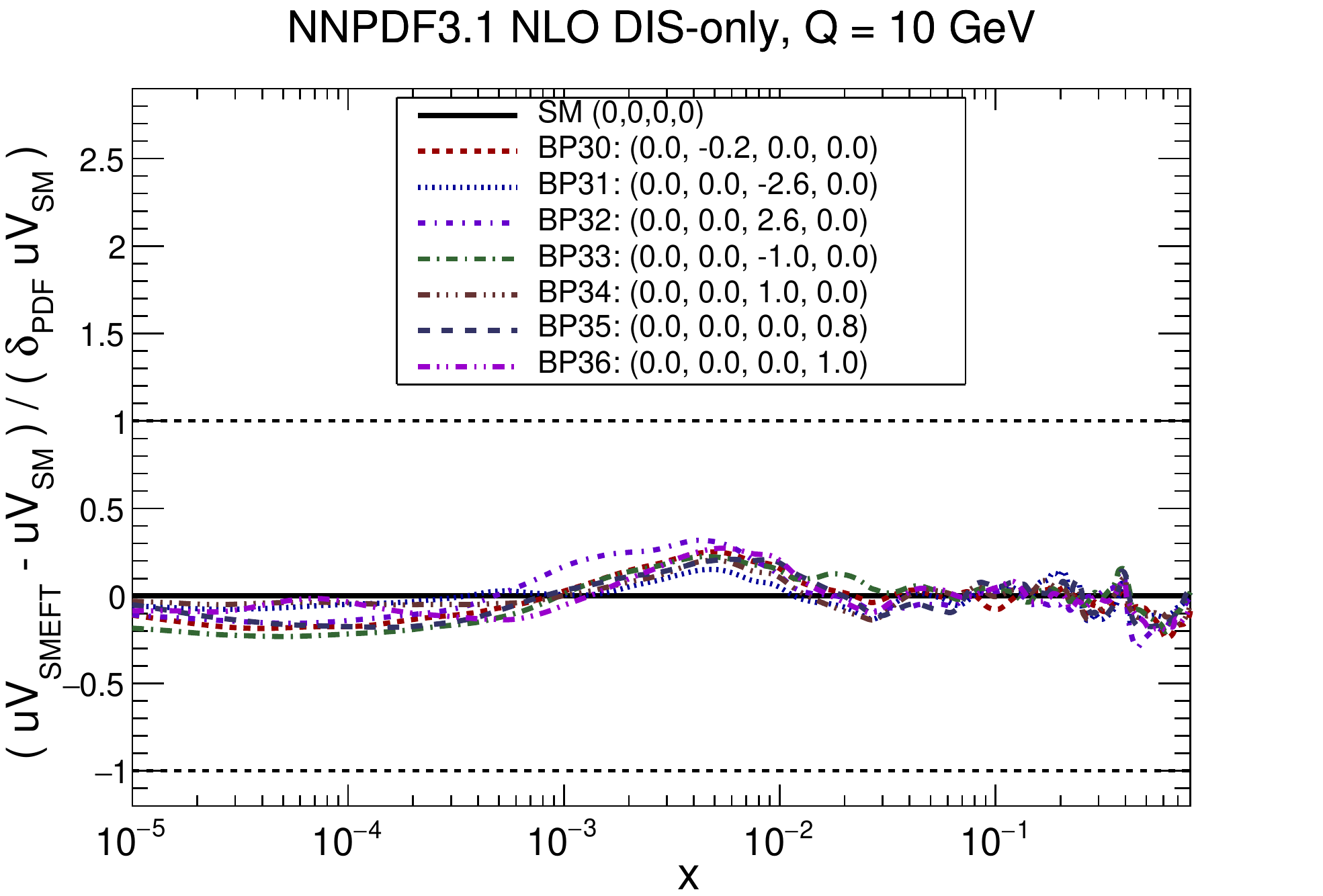}
    \includegraphics[width=0.49\textwidth]{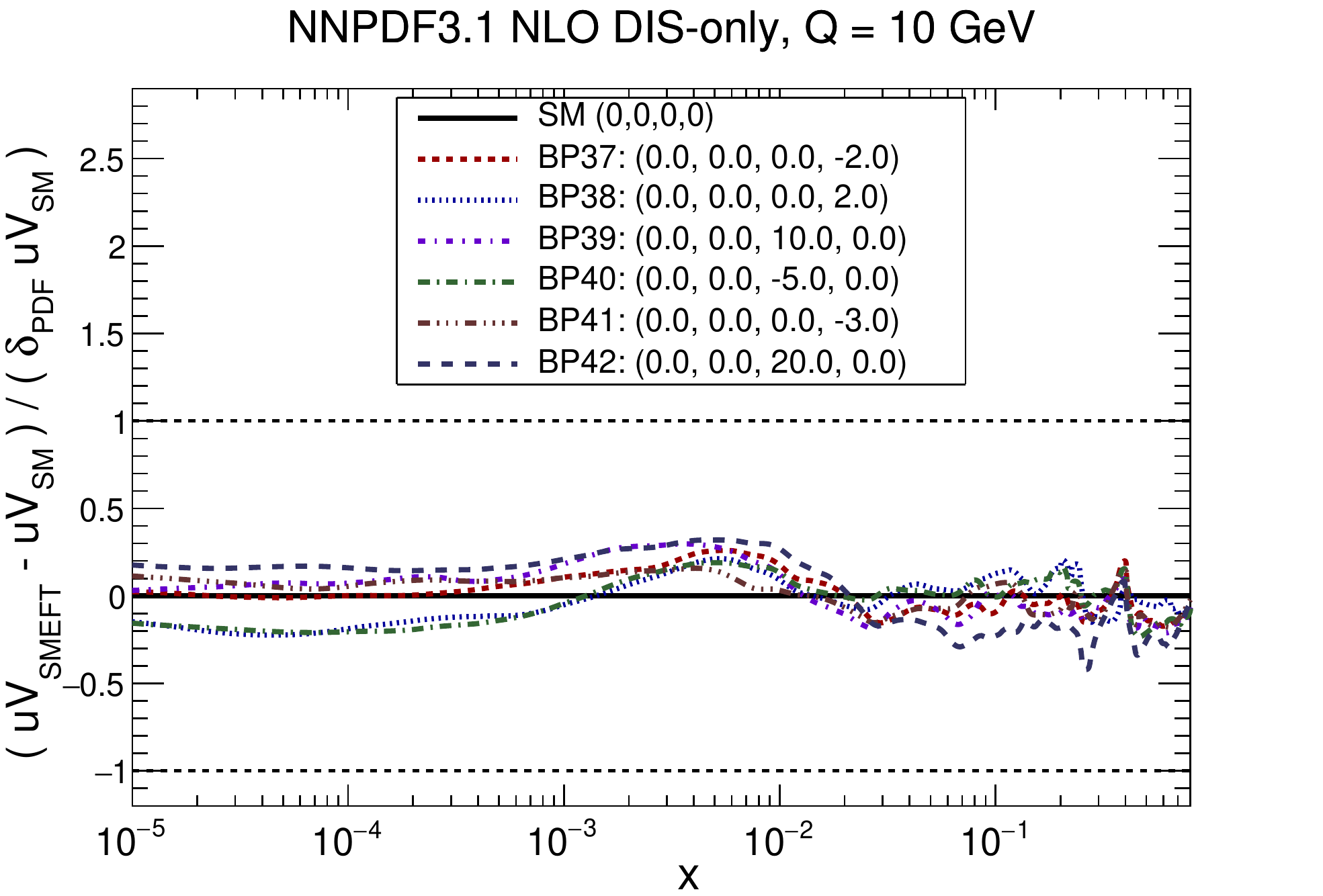}
  \end{center}
  \vspace{-0.3cm}
  \caption{Same as Fig.~\ref{fig:smeft-pull}, now for the up valence
    quark PDF $u_V(x,Q^2)$.
  }
  \label{fig:smeft-pull-2}
\end{figure}

In Figs.~\ref{fig:smeft-pull} and~\ref{fig:smeft-pull-2} we show these
pulls for the gluon 
and up valence quark PDFs  at $Q=10$ GeV for
all BPs listed in Table~\ref{tab:listfits}.
The results for other relevant quark combinations, such as the down valence
$d_V$, are similar to those for $u_V$ and thus they are not shown explicitly here.
In the case of the gluon, one finds that the the overall
magnitude of the SMEFT-induced distortion varies with the specific BP.
The values of the pulls are contained within the PDF uncertainty bands
for all BPs considered here, though for some BPs the value of the pull reaches
$P\simeq 0.5$ or even larger.
As expected, these pulls are close to zero for $x\lsim 10^{-3}$, since the small-$x$
region is mostly uncorrelated with the large-$Q^2$ one where the SMEFT corrections
are the largest (see Fig.~\ref{fig:SMEFT-corrections-BP}).

The situation is somewhat different for the up valence
quark distribution, Fig.~\ref{fig:smeft-pull-2}.
In this case, as for the other quark distributions, we find that the differences
between the fits based on  SM and the SMEFT-augmented calculations are smaller
than for the gluon, with $|P_{u_V}|\lsim 0.4$ for all BPs under consideration.
This result can be explained by noting that, within a DIS-only fit, the large-$x$ quarks
are constrained mostly from the fixed-target structure function data from the SLAC,
NMC, and BCDMS experiments.
The kinematic coverage of these fixed-target scattering measurements is restricted
to the region $Q^2 \lsim 300$ GeV$^2$, where SMEFT effects are strongly suppressed.
Therefore, large-$x$ quark PDFs are fixed by measurements
in a region where SMEFT effects are always small.

\begin{figure}[t]
  \begin{center}
    \includegraphics[width=0.49\textwidth]{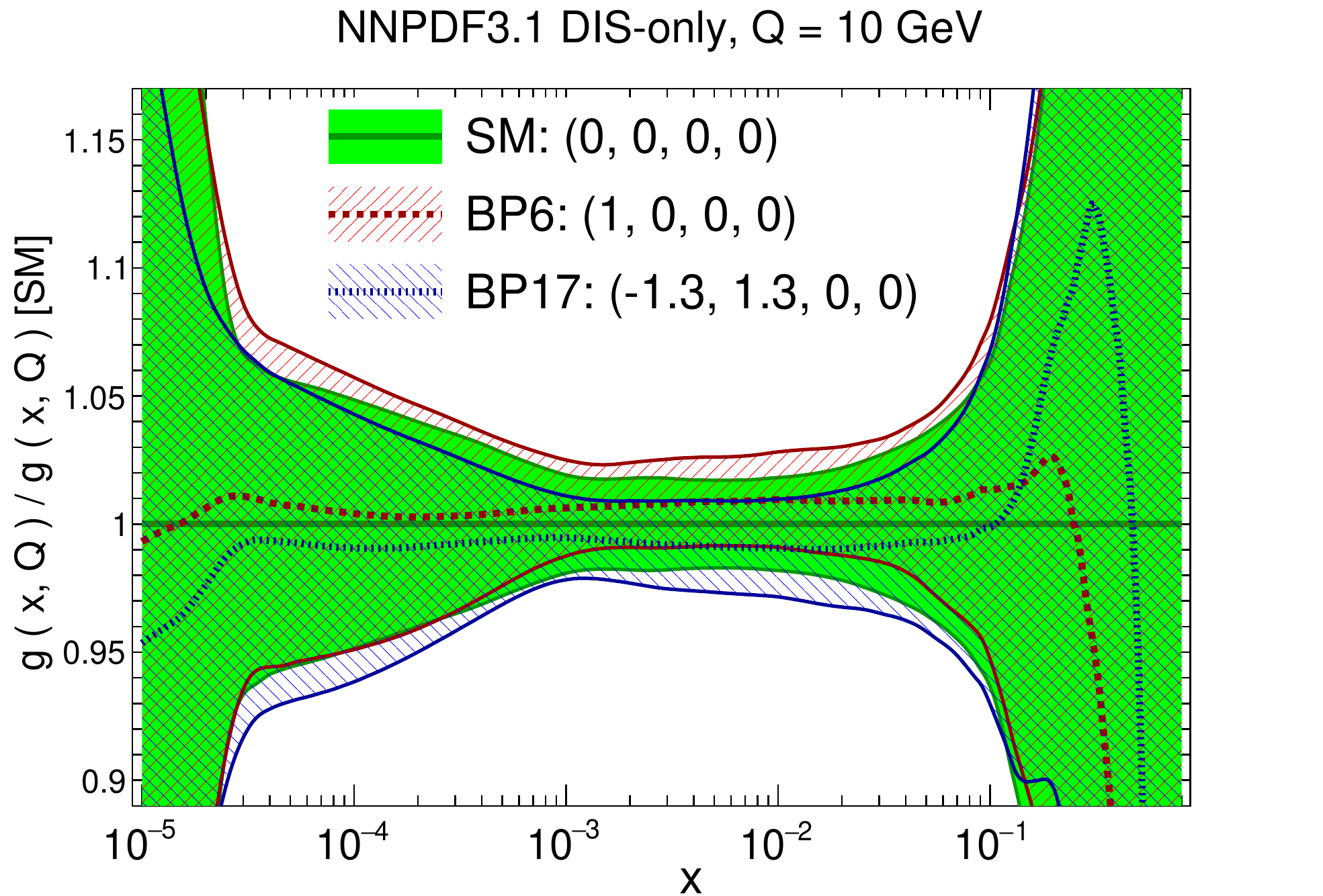}
    \includegraphics[width=0.49\textwidth]{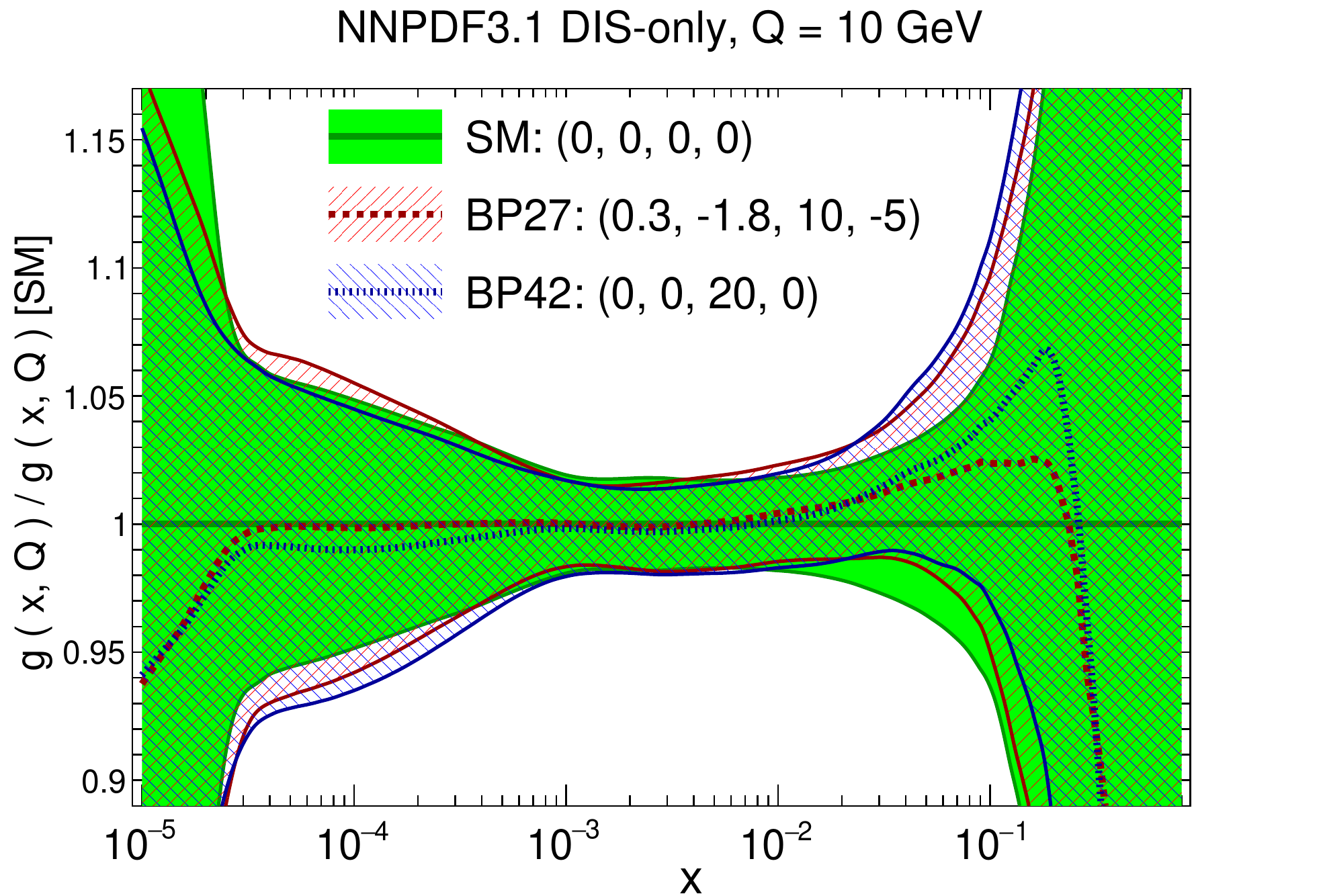}
    \includegraphics[width=0.49\textwidth]{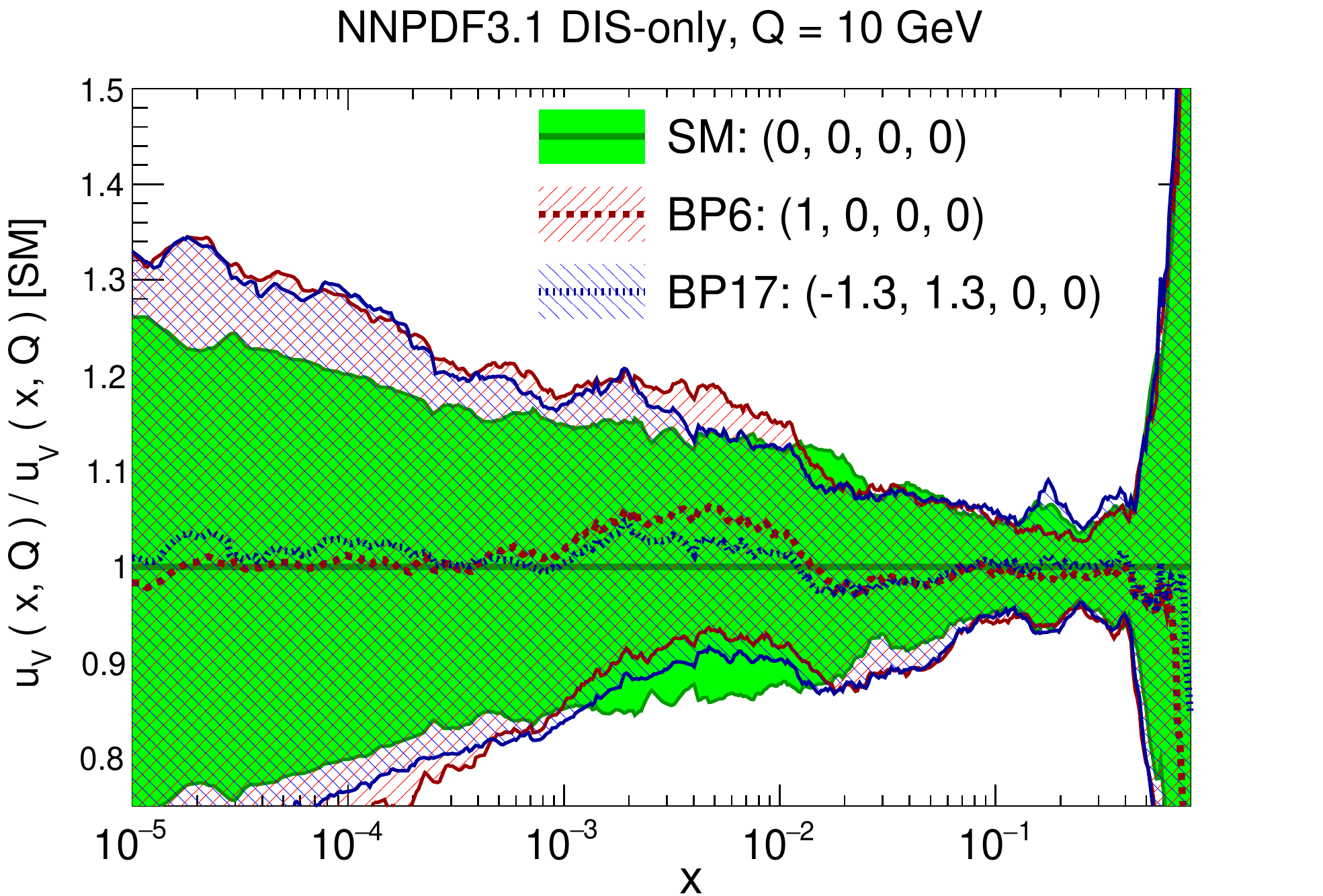}
    \includegraphics[width=0.49\textwidth]{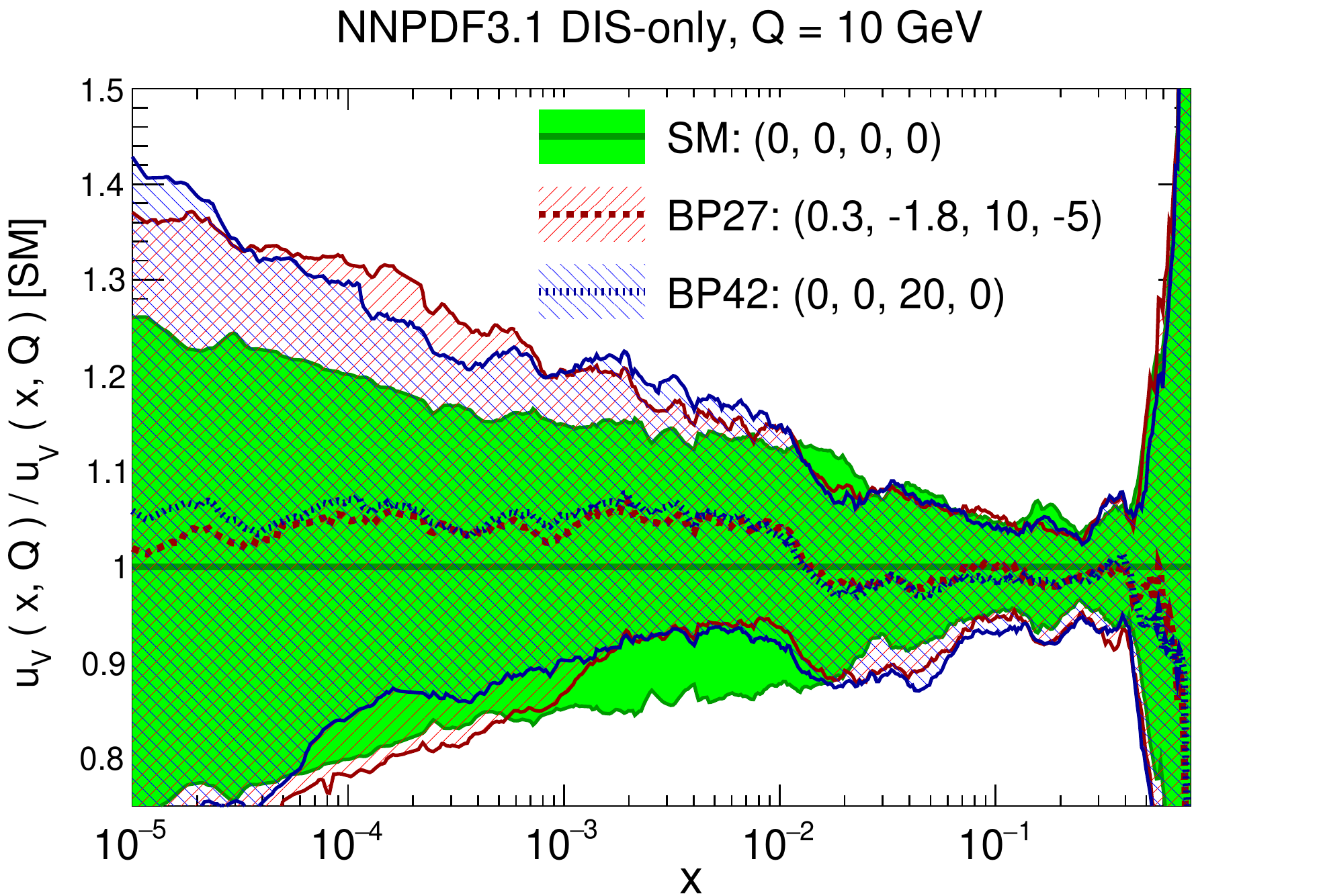}
  \end{center}
  \vspace{-0.3cm}
  \caption{The gluon (upper) and up valence quark (lower plots) at $Q=10$ GeV
    with the corresponding one-sigma uncertainties,
    where the PDF fits corresponding to representative SMEFT benchmark points
    are compared to the SM result.
  }
  \label{fig:smeft-fits-bp}
\end{figure}

In Fig.~\ref{fig:smeft-fits-bp} we display
the gluon  and up valence quark PDFs at $Q=10$ GeV 
comparing the results obtained using SM theory with those obtained
with representative SMEFT BPs, normalised to the
central value of the former.
We have chosen some of the BPs from Table~\ref{tab:listfits} for which
the PDFs differ the most as compared to the SM result, as quantified
by the pulls in Figs.~\ref{fig:smeft-pull} and~\ref{fig:smeft-pull-2}.
In particular, we show in this comparison BP6, BP17, BP27, and BP42.
In these comparison, the PDF uncertainties correspond
to the the 68\% CL intervals computed
from the corresponding $N_{\rm rep}=100$ replicas.
From Fig.~\ref{fig:smeft-fits-bp} we can observe that
these PDF comparisons are consistent with those of the corresponding pulls.
In particular, we note how the SMEFT-induced distortions are smaller
for the up valence quark than for the gluon, and how in the latter case
these effects are clearly visible especially at medium
and large-$x$.

Interestingly, not all the BPs shown
in Fig.~\ref{fig:smeft-fits-bp} lead to large values for $\Delta\chi^2_{\rm smeft}$.
As one can read from Table~\ref{tab:listfits}, while at
the post-fit level one has $\Delta\chi^2_{\rm smeft}=95$
and 64 for BP17 and BP42 respectively, instead one obtains
rather smaller differences, $\Delta\chi^2_{\rm smeft}=22$
and 14, for BP6 and BP27.
This result highlights that the SMEFT effects can induce non-negligible modifications
into the PDFs even in those cases where the overall fit quality is similar to that
of the SM case.

Additional information to understand the results of the PDF fits
with SMEFT corrections presented in this work
is provided by the  correlation pattern between the PDFs and the SMEFT
coefficients $\{a_q\}$.
These correlation coefficients can be evaluated,
for example in the case of the gluon, as follows
\be
\label{eq:correlationL2CT}
\rho\lp a_q, g(x,Q)\rp=\frac{\frac{1}{N_{\rm rep}}\sum_{k=1}^{N_{\rm rep}}
a_q^{\rm (min)(k)}g^{(k)}(x,Q) - \la a^{\rm (min)}_q\ra \la g(x,Q)\ra
}{\sqrt{\la a_q^{\rm (min)2}\ra - \la a_q^{\rm (min)}\ra^2}\sqrt{\la g(x,Q)^2\ra - \la g(x,Q)\ra^2} } \, ,
\ee
where $a_q^{\rm (min)(k)}$ stands for the minimum of the one-dimensional parabolic fit for
the coefficient of the four-fermion operator $\mathcal{O}_{lq}$, obtained using as input
for the theory calculation the $k$-th replica of NNPDF3.1 NNLO DIS-only.
In Eq.~(\ref{eq:correlationL2CT}), averages are computed over the $N_{\rm rep}=100$ replicas.
This correlation coefficient provides a measure of how the variations
in the input PDFs replica by replica translate into modifications of the best fit value
of the Wilson coefficient $a_q^{\rm (min)(k)}$ when the other coefficients are being set
to their SM values.

\begin{figure}[t]
  \begin{center}
    \includegraphics[width=0.49\textwidth]{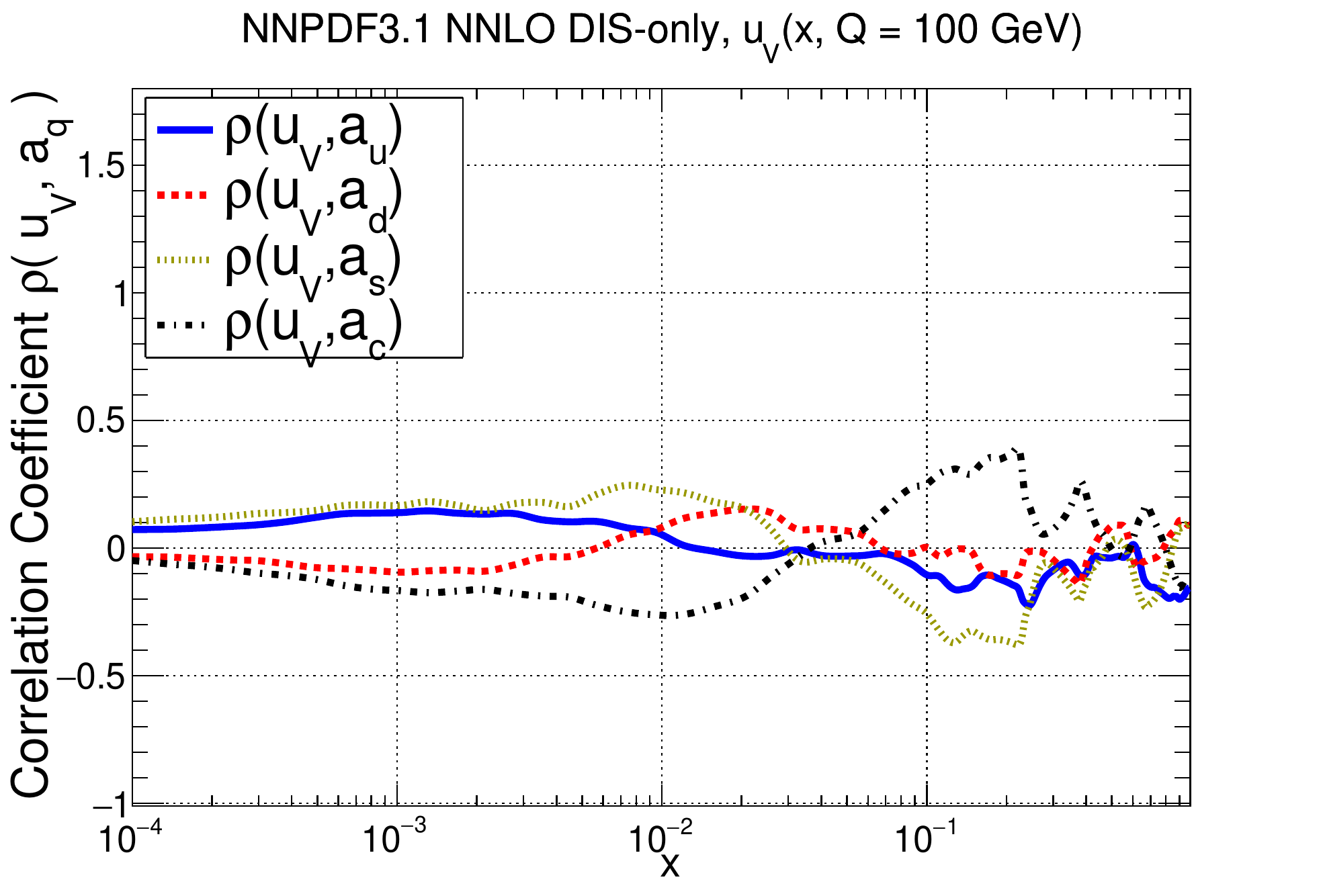}
    \includegraphics[width=0.49\textwidth]{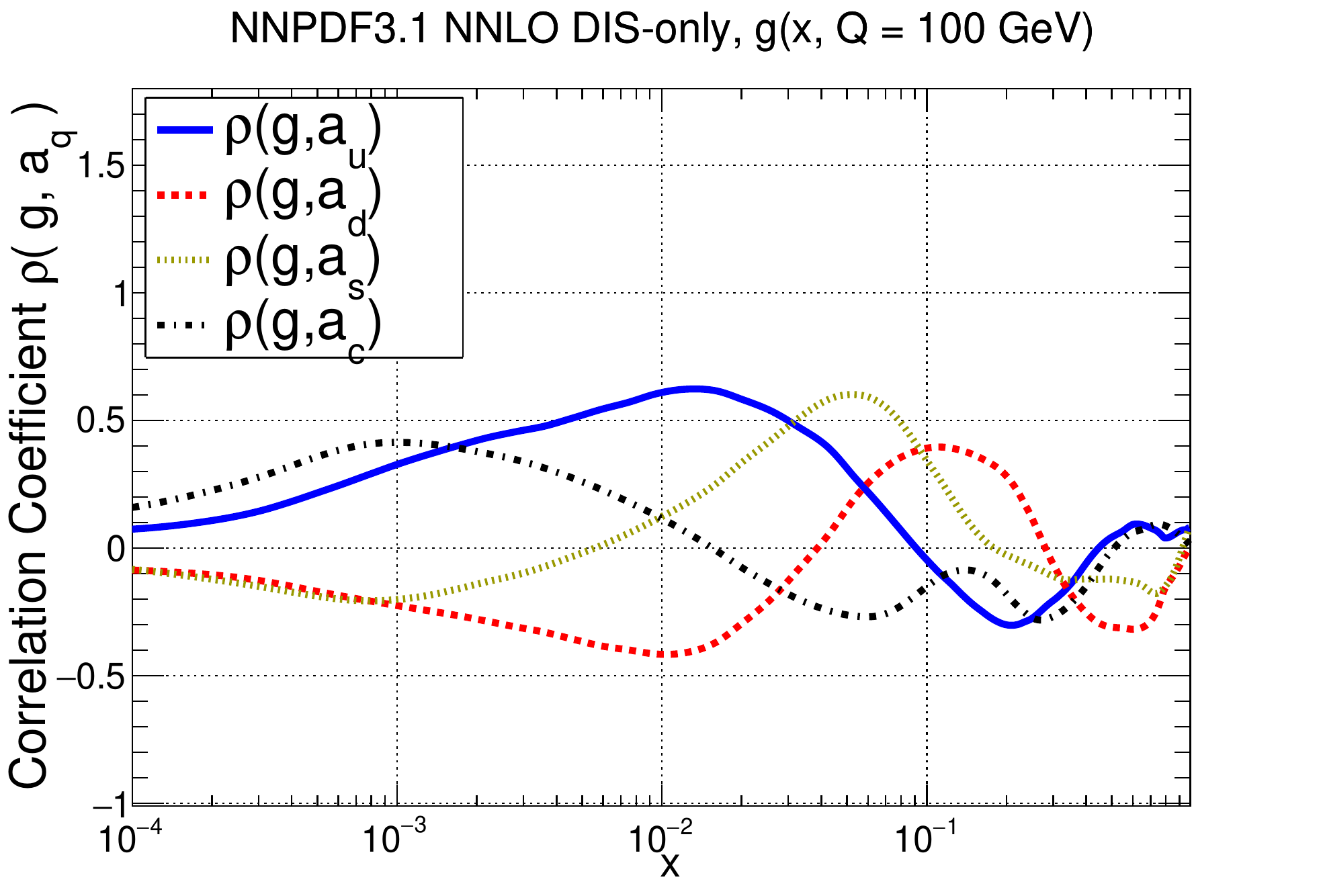}
  \end{center}
  \vspace{-0.4cm}
  \caption{The correlation coefficients $\rho$, Eq.~(\ref{eq:correlationL2CT}),
    in the NNPDF3.1 NNLO DIS-only fits at $Q=100$ GeV
    between the PDFs
    and the SMEFT degrees of freedom $a_u$, $a_d$, $a_s$, $a_c$.
    We show results for the gluon (left) and the up valence quark (right) PDFs.
  }
  \label{fig:correlations}
\end{figure}

In Fig.~\ref{fig:correlations} we show these correlation coefficients
between the up valence quark and the gluon PDFs in 
NNPDF3.1 NNLO DIS-only fits at $Q=100$ GeV.
Each of the curves corresponds to one of four
SMEFT coefficients: $a_u$, $a_d$, $a_s$, or $a_c$.
In the case of the up quark, the correlation with the SMEFT coefficients is weak
(with $|\rho|<0.4$ in all cases), consistent with the moderate impact that using
SMEFT-augmented calculations have on the corresponding PDF fits.
The correlations are more important for the gluon in the intermediate-$x$ region, reaching $\rho \simeq 0.6$, again mirroring the corresponding
results at the PDF fit level.
The fact that the correlations have a similar shape but opposite overall sign
for $a_u$ and $a_d$ is a consequence of the fact that their contributions
at the structure function level have opposite signs.
Note that these correlation patterns are specific of the input dataset and choice of SMEFT
operator basis, and will in general be rather different if either of the two were to be varied.

%% file: chi2-smeft-nnpdf31dis.tex
\begin{table}[p]
  \centering
  \footnotesize
  \renewcommand{\arraystretch}{1.50}
   \begin{tabular}{|C{1.1cm}|C{1.0cm}C{1.0cm}C{1.0cm}C{1.0cm}||C{1.6cm}|C{1.6cm}|C{1.6cm}||C{1.6cm}|C{1.6cm}|C{1.6cm}|}
     \hline
     &   \multirow{2}{*}{\bf $a_u$}     &  \multirow{2}{*}{\bf $a_d$}  &
     \multirow{2}{*}{\bf $a_s$}  &  \multirow{2}{*}{\bf $a_c$}  & \multicolumn{3}{c||}{{\bf Pre-fit} (fixed input PDF)}
      & \multicolumn{3}{c|}{{\bf Post-fit} }\\
       &        &    &    &     & $\chi^2_{\rm tot}$
      & $\chi^2_{\rm hera}$ & $\Delta \chi^2_{\rm smeft}$  & $\chi^2_{\rm tot}$
      & $\chi^2_{\rm hera}$ & $\Delta \chi^2_{\rm smeft}$ \\
       \hline
 {\bf SM}&     0  &   0   &  0   &  0   & 3445.8      &  1311.8  & -  & 3445.8      &  1311.8  & -   \\
      \hline
BP1  &  -0.28  &	0.1 &	0.1 &	-0.28		 &	3453.4 &	1319.4	& 7.6 &	3451.0 &	1314.6	& 5.2 \\
BP2  &  -0.04 &	-0.19 &	-0.19&	-0.04	 	&	3445.8	& 1311.7 &	0.0	&	3447.2	& 1312.4 &	1.4 \\
BP3  &	 -1.0&	0.7  &	-0.7 &	1.0	 	&	3502.8 &	1368.9 &	57.0
&	3490.2 &	1354.9 &	44.4 \\
BP4  &	-0.7 &	0.5  &	0.0  &	3.0		 	&	3473.0	& 1338.7 &	27.2
	&	3470.7	& 1331.1 &	24.9 \\
BP5  &	1.0  &	0.0  &	0.0  &	0.0		 	&	3474.1	& 1339.9 &	28.3
	&	3465.5	& 1341.7 &	19.7\\
BP6  &	-0.5 &	0.0  &	0.0  &	0.0		 	&	3461.1	& 1327.1 &	15.3
	&	3468.3	& 1324.1 &	22.5\\
BP7  &	0.5 &	0.0  &	0.0  &	0.0		 	&	3450.2	& 1316.1&	4.4
	&	3453.7	& 1316.9&	7.9\\
BP8  &	0.3 &	0.0  &	0.0  &	0.0		 	&	3446.1	& 1312.0 &	0.3
	&	3440.2	& 1313.7 &	-5.6\\
BP9  &	0.0 &	-1.0  &	0.0  &	0.0		 	&	3445.5	& 1311.4 &	-0.3
	&	3443.2	& 1312.5 &	-2.6\\
BP10  & 0.0 &	0.5  &	0.0  &	0.0	 	&	3448.1	& 1314.1 &	2.3
	&	3440.7	& 1315.4 &	-5.1 \\
BP11 &	0.0 &	-1.5  &	0.0  &	0.0	 	&	3447.3 & 	1313.2 &	1.5
&	3442.5 & 	1318.9 &	-3.2\\ 
BP12 &	0.0 &	-1.9  &	0.0  &	0.0	 	&	3449.8	& 1315.6 &	4.0
	&	3448.9	& 1317.4 &	 3.1\\
BP13 &	0.0 &	0.0  &	-0.7  &	0.0		&	3446.2	& 1312.1 &	0.4
&	3440.4	& 1312.5 &	-5.4 \\
BP14 &	0.0 &	0.0  &	0.0  &	-0.2	 	&	3445.7	& 1311.6 &	-0.1
	&	3436.5	& 1314.2 &	-9.3\\
BP15 &	0.0 &	0.0  &	0.0  &	-1.0	 	&	3445.5	& 1311.6 &	-0.3
&	3444.0	& 1311.3 &	-1.8\\
BP16 &	0.9 &	0.9  &	0.0  &	0.0		 	&	3457.5 &	1323.4 &	11.7
	&	3455.2 &	1325.5 &	9.4\\
BP17  &  -1.3 &	1.3  &	0.0  &	0.0 	&	3566.8	& 1433.0 &	121.0
&	3541.1	& 1405.7 &	95.3\\
BP18  &	0.0  &	0.0  & 	5.0  &	-5.0		 	&	3481.6 &	1347.8 &	35.8
	&	3470.1 &	1337.6 &	24.4\\
BP19 &	0.0  &	0.0  &	-2.0  &	2.0		 	&	3455.4	& 1321.2 &	9.6
	&	3448.9	& 1323.4 &	3.1\\
BP20 &	0.3  & 	0.0  &	10.0  &	0.0	 	&	3454.6	& 1320.6 &	8.8
	&	3451.4	& 1321.6 &	-1.8\\
BP21 &	0.3  & 	0.0  &	-5.0  &	0.0	 	&	3458.1	& 1324.0 &	12.3
	&	3454.2	& 1327.0 &	8.4\\
BP22 &	-0.3  & 	0.0  &	0.0  &	5.0	 	&	3465.2	& 1330.6 &	19.4
	&	3463.2	& 1326.7 &	17.4\\
BP23 &	0.0  & 	1.2  &	10.0  &	0.0	 	&	3487.1	& 1353.2 &	41.3
	&	3474.7	& 1343.1 &	28.9\\
BP24 &	0.0  & 	-1.8  &	-5.0&	0.0&	3467.2	& 1332.9 &	21.4
	&	3469.8	& 1336.3 &	24.0\\
BP25 &	0.3  & 	1.2  &	-5.0&	-5.0&	3449.3	& 1315.4 &	3.5
	&	3436.9	& 1311.3 &	-8.9\\
BP26 &	0.3  & 	-1.8  &	10.0&	5.0&	3466.7	& 1332.0 &	20.9
	&	3456.6	& 1327.1 &	10.8\\
BP27 &	0.3  & 	-1.8  &	10.0&	-5.0&	3462.7	& 1328.8 &	16.9
	&	3459.7	& 1324.5 &	13.9\\
BP28 &	0.3  & 	-1.8  &	-5.0&	-5.0&	3445.1	& 1311.0 &	-0.7
	&	3437.1	& 1310.9 &	-8.9\\
BP29 &	-0.3  & 	-1.8  &	-5.0&	5.0&	3510.7	& 1375.9 &	64.9
	&	3493.4	& 1365.7 &	47.6\\
BP30 &	0.0  & 	-0.2  &	0.0&	0.0&	3445.3	& 1311.2 &	-0.5
	&	3437.8	& 1308.3 &	-8.0\\
BP31 &	0.0  & 	0.0  &	-2.6&	0.0&	3448.1	& 1314.0 &	2.3
	&	3446.4	& 1313.2 &	0.6\\
BP32 &	0.0  & 	0.0  &	2.6&	0.0&	3446.6	& 1312.5 &	0.8
	&	3443.3	& 1307.8 &	-2.5\\
BP33 &	0.0  & 	0.0  &	-1.0&	0.0&	3446.4	& 1312.3 &	0.6
	&	3442.0	& 1313.9 &	-3.8\\
BP34 &	0.0  & 	0.0  &	1.0&	0.0&	3445.8	& 1311.7 &	0.0
	&	3439.1	& 1312.0 &	-6.7\\
BP35 &	0.0  & 	0.0  &	0.0&	0.8&	3447.2	& 1313.1 &	1.4
	&	3449.5	& 1314.3 &	3.7\\
BP36 &	0.0  & 	0.0  &	0.0&	1.0&	3447.7	& 1313.6 &	1.9
	&	3442.8	& 1315.7 &	-3.0\\
BP37 &	0.0  & 	0.0  &	0.0&	-2.0&	3446.7	& 1312.8 &	0.9
	&	3444.1	& 1312.2 &	-1.7\\
BP38 &	0.0  & 	0.0  &	0.0&	2.0&	3451.0	& 1316.8 &	5.2
	&	3450.6	& 1317.0 &	4.8\\
BP39 &	0.0  & 	0.0  &	10.0&	0.0&	3464.5	& 1330.6 &	18.7
	&	3461.0	& 1327.4 &	15.2\\
BP40 &	0.0  & 	0.0  &	-5.0&	0.0&	3452.8	& 1318.7 &	7.0
	&	3444.0	& 1317.4 &	-1.8\\
BP41 &	0.0  & 	0.0  &	0.0&	-3.0&	3449.5	& 1315.6 &	3.7
	&	3446.1	& 1314.9 &	0.3\\
BP42 &	0.0  & 	0.0  &	20.0&	0.0&	3526.6	& 1392.7 &	80.8
	&	3509.9	& 1368.9 &	64.1\\
\hline  \end{tabular}
   \caption{ The SMEFT  benchmark points (BPs)
     for which variants of the NNPDF3.1 NNLO DIS-only fit have been produced.
     For each BP, we indicate the values
     of the total and HERA-only $\chi^2$, denoted by $\chi^2_{\rm tot}$
     and $\chi^2_{\rm hera}$ respectively, both in the case of a common fixed input
     PDF set (``pre-fit'' column) and once the PDF has been fitted to the theory
     based on the corresponding BP (``post-fit'' column).
     We also indicate the absolute difference
     between $\chi^2_{\rm tot}$ in the SMEFT and in the SM, 
     $\Delta \chi^2_{\rm smeft}$.
     Note that throughout this work we are assuming that $\Lambda=1$ TeV.
     \label{tab:listfits}
  }
\end{table}